\documentclass[12pt]{article}
\makeatletter
\@addtoreset{equation}{section}
\makeatother

\renewcommand{\title}[1]{\null\vspace{25mm}

\noindent{\Large{\bf #1}}\vspace{10mm}

\noindent {\large By }

}
\newcommand{\authors}[1]{\noindent{\large #1}\vspace{3mm}

}
\newcommand{\address}[1]{\noindent #1\vspace{5mm}

}
\renewcommand{\abstract}[1]{\vspace{9mm}

\noindent{\small{\em Abstract.} #1}\vspace{2mm}
}

\usepackage{amssymb}
\usepackage{latexsym}
\usepackage{amsthm}
\theoremstyle{plain}
\newtheorem{theorem}{Theorem}[section]
\newtheorem{definition}[theorem]{Definition}
\newtheorem{lemma}[theorem]{Lemma}

\newcommand{\uline}{\vrule height.06ex depth.02ex width.6em}
\newcommand{\mvee}{\vee\kern-.69em\uline}

\font\1=cmss10

\begin{document}

\markright{\it International Journal of Theoretical Physics\/\rm, 
\bf 39\rm, 2337--2379 (2000).}

\font\1=cmss10

\title{Equations and State and Lattice Properties That Hold\\[2mm]
in Infinite Dimensional Hilbert Space}

\authors{Norman D.~Megill$^\dag$$\footnote{E-mail: nm@alum.mit.edu;
Web page: http://www.shore.net/\~{}ndm/java/mm.html}$
and Mladen Pavi\v ci\'c$^\ddag$$\footnote{E-mail: mpavicic@faust.irb.hr;
Web page: http://m3k.grad.hr/pavicic}$}
\address{$^\dag$Boston Information Group, 30 Church St.,
Belmont MA 02478, U.~S.~A.\\
$^\ddag$Dept.~of Physics, University of Maryland, Baltimore County,
Baltimore, MD 21250, U.~S.~A.\\
and University of Zagreb, Gradjevinski Fakultet, Ka\v ci\'ceva 26, HR-10000
Zagreb, Croatia.}
\abstract{We provide several new results on quantum state space, on
lattice of subspaces of an infinite dimensional Hilbert space, and on
infinite dimensional Hilbert space equations as well as on connections
between them. In particular we obtain an $n$-variable generalized
orthoarguesian equation which holds in any infinite dimensional
Hilbert space. Then we strengthen Godowski's result by
showing that in an ortholattice on which strong states are defined
Godowski's equations as well as the orthomodularity hold. We also
prove that all 6- and 4-variable orthoarguesian equations
presented in the literature can be reduced to new 4- and
3-variable ones, respectively and that Mayet's examples follow
from Godowski's equations. To make a breakthrough in testing these
massive equations we designed several novel algorithms for generating
Greechie diagrams with an arbitrary number of blocks and atoms
(currently testing with up to 50) and for automated checking of
equations on them. A way of obtaining complex infinite dimensional
Hilbert space from the Hilbert lattice equipped with several
additional conditions and without invoking the notion of state is
presented. Possible repercussions of the results to quantum computing
problems are discussed.}



\vbox to 2cm{\vfill}

\section{Introduction}
\label{sec:intro}

\markright{Megill and Pavi\v ci\'c (2000)}

Recent theoretical and experimental developments in the field of
quantum computing opened the possibility of using quantum mechanical
states, their superpositions, and operators defined on them---i.e.,
the Hilbert space formalism---to exponentially speed up
computation of various systems, on the one hand, and to simulate
quantum systems, on the other. Quantum computers can be looked upon
as parallel computing machines. Looking at the speed of computation,
the difference between classical and quantum parallel machines is
that in a classical one we increase its speed by increasing its
physical space (occupied by electronic components: processors, etc.)
while in a quantum one we achieve this by exponentially increasing
its state space by means of linearly increased physical space
(a register of $n$ quantum bits---{\em qubits}---prepares
a superposition of $2^n$ states). To make a quantum parallel machine
compute a particular problem is tricky and requires a great deal of
ingenuity but a computed system itself need not be quantum and
need not be simulated. Actually, all algorithms designed so
far are of such a kind. For example, Shor's algorithm \cite{shor}
factors $n$-digit numbers, Grover's algorithm \cite{grover} searches
huge databases, and Boghosian-Taylor's algorithm \cite{schr-simul}
computes the Schr\"odinger equation. As opposed to this, a quantum
simulator would simulate a quantum system (e.g., an atom,
a molecule,\dots) and give its final state directly: a quantum
computer working as a quantum simulator would not solve the
Scr\"odinger equation but would simulate it and the outputs
would be its solutions---by typing in a Hamiltonian at a console
we would simulate the system.

Quantum simulation of quantum systems describable in the Hilbert
space formalism (by the Scr\"odinger equation) would however only
then be possible if we found an algebra underlying Hilbert
space in the same way in which the Boolean algebra underlies classical
state space. Such an algebra for quantum computers has recently been
named {\em quantum logic} in some analogy to classical logic
for classical computer. \cite{plenio} However, this name is misleading
for both types of computers because  proper logics, both classical
and quantum, have at least two models each. \cite{mpcommp99} Classical
logic has not only a Boolean algebra but also a non-orthomodular
algebra as its model and quantum logic not only an orthomodular
algebra (Hilbert space) but also another non-orthomodular algebra:
a weakly orthomodular lattice. What resolves this ambiguity is that
as soon as we require either a numerical or a probabilistic evaluation
of the propositions of classical logic we are left only with the
Boolean algebra \cite{mpcommp99} and that as soon as we impose
probabilistic evaluation (states) on quantum logic we are
left only with Hilbert space. Therefore quantum logic itself does
not to play a role in the current description of quantum systems. Its
standard model---Hilbert space---does.

One can make Hilbert space operational on a quantum computer
by imposing lattice equations that hold in any Hilbert space
on the computer states using quantum gates. Unfortunately not very
much is known about the equations---explorations of Hilbert
space have so far been concentrated on the operator theory
leaving the theory of the subspaces of a Hilbert space (wherefrom
we obtain these equations) virtually unexplored. So, in this paper
we investigate how one can arrive at such equations starting
from both algebraic and probabilistic structures of
Hilbert space of quantum measurement and computation. We obtain
several new results on these structures, give a number of new
Hilbert space equations, and systematize, significantly simplify,
and mutually reduce already known equations.
In Section \ref{sec:oml-eqs} we give several new characterizations
of orthomodularity which we make use of later on.
In Section \ref{sec:states} we consider ways in which states can
be defined on an ortholattice underlying Hilbert space
and make it orthomodular---when quantum, and distributive---when
classical. We also analyze several kinds of equations characteristic
of strong states in Hilbert space (Godowski's and Mayet's).
On the other hand we present a way of obtaining complex Hilbert
space from the Hilbert lattice equipped with several additional
conditions but without invoking the notion of state.
In Section \ref{sec:oa} we give a new way of presenting
orthoarguesian equations and their consequences
which must hold in any Hilbert space and for which it is not known
whether they are characteristic of the states or not. We reduce the
number of variables used in the orthoarguesian-like equations in the
literature (from 6 to 4 for the standard orthoarguesian equation and
to 3 for all its consequences), we show that all consequences that
appear in the literature reduce to a single 3-variable equation, and
find a new one which does not. In Section \ref{sec:5oa} we present
generalizations of orthoarguesian equations that must hold in any
Hilbert space. This previously unknown and unconjectured result
is our most important contribution to the theory of infinite
dimensional Hilbert spaces in this paper.
In Section \ref{sec:distr} we show several distributive
properties that must hold in any Hilbert space.

\section{\large Orthomodular Lattice Underlying Hilbert Space }
\label{sec:oml-eqs}

Closed subspaces of Hilbert space form an algebra called a Hilbert
lattice.  A Hilbert lattice is a kind of orthomodular lattice which we,
in this section, introduce starting with an ortholattice which is a
still simpler structure.  In any Hilbert lattice
the operation \it meet\/\rm, $a\cap b$, corresponds to
set intersection, ${\cal H}_a\bigcap{\cal H}_b$, of subspaces ${\cal
H}_a,{\cal H}_b$ of Hilbert space ${\cal H}$, the ordering relation
$a\le b$ corresponds to ${\cal H}_a\subseteq{\cal H}_b$, the operation
\it join\/\rm, $a\cup b$, corresponds to the smallest closed subspace of
$\cal H$ containing ${\cal H}_a\bigcup{\cal H}_b$, and $a'$ corresponds
to ${\cal H}_a^\perp$, the set of vectors orthogonal to all vectors in
${\cal H}_a$. Within Hilbert space there is also an operation which
has no a parallel in the Hilbert lattice: the sum of two subspaces
${\cal H}_a+{\cal H}_b$ which is defined as the set of sums of vectors
from ${\cal H}_a$ and ${\cal H}_b$. We also have
${\cal H}_a+{\cal H}_a^\perp={\cal H}$. One can define
all the lattice operations on Hilbert space itself following the above
definitions (${\cal H}_a\cap{\cal H}_b={\cal H}_a\bigcap{\cal H}_b$,
etc.). Thus we have
${\cal H}_a\cup{\cal H}_b=\overline{{\cal H}_a+{\cal H}_b}=
({\cal H}_a+{\cal H}_b)^{\perp\perp}=
({\cal H}_a^\perp\bigcap{\cal
H}_b^\perp)^\perp$,\cite[p.~175]{isham} where
$\overline{{\cal H}_c}$ is a closure of ${\cal H}_c$, and therefore
${\cal H}_a+{\cal H}_b\subseteq{\cal H}_a\cup{\cal H}_b$.
When ${\cal H}$ is finite dimensional or when
the closed subspaces ${\cal H}_a$ and  ${\cal H}_b$ are orthogonal
to each other then ${\cal H}_a+{\cal H}_b={\cal H}_a\cup{\cal H}_b$.
\cite[pp.~21-29]{halmos}, \cite[pp.~66,67]{kalmb83},
\cite[pp.~8-16]{mittelstaedt-book}

The projection associated with ${\cal H}_a$ is given by
$P_a(x)=y$ for vector $x$ from ${\cal H}$ that has a unique
decomposition $x=y+z$ for $y$ from ${\cal H}_a$ and $z$ from
${\cal H}_a^\perp$. The closed subspace belonging to $P$ is
${\cal H}_P=\{x\in {\cal H}|P(x)=x\}$. Let $P_a\cap P_b$ denote
a projection on  ${\cal H}_a\cap{\cal H}_b$, $P_a\cup P_b\>$
a projection on  ${\cal H}_a\cup{\cal H}_b$, $P_a+P_b\>$
a projection on  ${\cal H}_a+{\cal H}_b\>$ if ${\cal H}_a\perp
{\cal H}_b$, and let $P_a\le P_b$ mean
${\cal H}_a\subseteq{\cal H}_b$. Thus $a\cap b$ corresponds to
$P_a\cap P_b=\lim_{n\to\infty}(P_aP_b)^n
$,\cite[p.~20]{mittelstaedt-book} $a'$ to $I-P_a$,
$a\cup b$ to $P_a\cup P_b=I-\lim_{n\to\infty}[(I-P_a)(I-P_b)]^n
$,\cite[p.~21]{mittelstaedt-book}  and $a\le b$ to  $P_a\le P_b$.
$a\le b$ also corresponds to either
$P_a=P_aP_b$ or to $P_a=P_bP_a$ or to $P_a-P_b=P_{a\cap b'}$.
Two projectors commute iff their
associated closed subspaces commute. This means that (see
Definition \ref{def:commut}) $a\cap(a'\cup b)\le b$
corresponds to $P_aP_b=P_bP_a$. In the latter case we have:
$P_a\cap P_b=P_aP_b$ and $P_a\cup P_b=P_a+P_b-P_aP_b$.
$a\perp b$, i.e., $P_a\perp P_b$ is characterized by $P_aP_b=0$.
\cite[pp.~173-176]{isham}, \cite[pp.~66,67]{kalmb83},
\cite[pp.~18-21]{mittelstaedt-book}, \cite[pp.~47-50]{holl70},

In this section we give several definitions of an orthomodular
lattice, two of which (given by Theorem \ref{th:other-eq}) are new.
In Section \ref{sec:states} we then show that the orthomodularity
of an ortholattice is a consequence of defining strong
states on the ortholattice and in Sections \ref{sec:states} and
\ref{sec:oa} we show that it also a consequence of other more
restrictive lattice conditions: Godowski equations and
orthoarguesian equations.

\begin{definition}
An ortholattice ({\em OL\/}) is an algebra
$\langle{\cal L}_{\rm O},',\cap,\cup\rangle$
such that the following conditions are satisfied for any
$a,b,c,d,e,f,g,h\in {\cal L_{\rm O}}$:
\begin{eqnarray}
(b\cap(c\cap a))\cup a &=& a\\
((a\cap(b\cap(f\cup c)))\cup d)\cup e &=& ((((g\cap g')
\cup(c'\cap f')')\cap(a\cap b))\cup e)\cup((h\cup d)\cap d)\ \ \ \ \
\end{eqnarray}
\end{definition}

\begin{lemma}\label{lemma:soboc}
The following conditions hold in any {\em OL\/}:
$a\cup b\>=\>b\cup a$, $(a\cup b)\cup c\>=\>a\cup (b\cup c)$,
$a''\>=\>a$, $a\cup (a\cap b)\>=\>a$, $a\cap b\>=\>(a'\cup b')'$.
Also an algebra in which these conditions
hold is an {\em OL}.
\end{lemma}
\begin{proof} As given in Ref.~\cite{soboc}.
\end{proof}

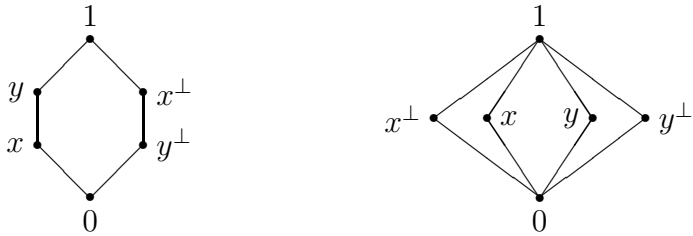
\begin{figure}[htbp]\centering
  \setlength{\unitlength}{1pt}
  \begin{picture}(240,90)(0,0)

    \put(30,0){
      \begin{picture}(60,80)(-10,-10)
        \put(20,0){\line(-1,1){20}}
        \put(20,0){\line(1,1){20}}
        \put(0,20){\line(0,1){20}}
        \put(40,20){\line(0,1){20}}
        \put(0,40){\line(1,1){20}}
        \put(40,40){\line(-1,1){20}}

        \put(20,-5){\makebox(0,0)[t]{$0$}}
        \put(-5,20){\makebox(0,0)[r]{$x$}}
        \put(45,20){\makebox(0,0)[l]{$y^\perp$}}
        \put(-5,40){\makebox(0,0)[r]{$y$}}
        \put(45,40){\makebox(0,0)[l]{$x^\perp$}}
        \put(20,65){\makebox(0,0)[b]{$1$}}

        \put(20,0){\circle*{3}}
        \put(0,20){\circle*{3}}
        \put(40,20){\circle*{3}}
        \put(0,40){\circle*{3}}
        \put(40,40){\circle*{3}}
        \put(20,60){\circle*{3}}

      \end{picture}
    } 

    \put(170,0){

      \begin{picture}(90,80)(-20,-10)
        \put(40,0){\line(-2,3){20}}
        \put(40,0){\line(2,3){20}}
        \put(40,0){\line(-4,3){40}}
        \put(40,0){\line(4,3){40}}
        \put(40,60){\line(-2,-3){20}}
        \put(40,60){\line(2,-3){20}}
        \put(40,60){\line(-4,-3){40}}
        \put(40,60){\line(4,-3){40}}

        \put(40,-5){\makebox(0,0)[t]{$0$}}
        \put(25,30){\makebox(0,0)[l]{$x$}}
        \put(85,30){\makebox(0,0)[l]{$y^\perp$}}
        \put(55,30){\makebox(0,0)[r]{$y$}}
        \put(-5,30){\makebox(0,0)[r]{$x^\perp$}}
        \put(40,65){\makebox(0,0)[b]{$1$}}

        \put(40,0){\circle*{3}}
        \put(0,30){\circle*{3}}
        \put(20,30){\circle*{3}}
        \put(60,30){\circle*{3}}
        \put(80,30){\circle*{3}}
        \put(40,60){\circle*{3}}
      \end{picture}
    } 

  \end{picture}
  \caption{\hbox to3mm{\hfill}(a)
    Lattice O6;
  \hbox to20mm{\hfill} (b) Lattice MO2.
\label{fig:o6mo2}}
\end{figure}

\begin{definition}\label{mixed-id}
An orthomodular lattice ({\em OML\/}) is an
ortholattice in which any one of the following hold
\begin{eqnarray}
a\equiv_i b=1\qquad \Rightarrow\qquad a=b, \qquad\qquad
i=1,\dots,5\label{eq:qm-as-id}
\end{eqnarray}
holds, where $a\equiv_i b\ {\buildrel\rm def\over =}\ (a\to_i b)\cap(b\to_0
a),\ i=1,\dots,5$,  where $a\to_0b\
{\buildrel\rm def\over =}\ a'\cup b$,\break $a\to_1b\
{\buildrel\rm def\over =}\ a'\cup(a\cap b)$,
$a\to_2b\ {\buildrel\rm
def\over =}\ b'\to_1a'$, $a\to_3b\
{\buildrel\rm def\over =}\ (a'\cap b)\cup(a'\cap b')\cup
(a\to_1b)$, $a\to_4b\ {\buildrel\rm def\over =}\
b'\to_3a'$, and $a\to_5b\ {\buildrel\rm def\over =}\
(a\cap b)\cup(a'\cap b)\cup(a'\cap b')$.
\end{definition}
The equivalence of this definition to the other definitions in the
literature follows from Lemma 2.1 and Theorem 2.2 of \cite{mpcommp99}
and the fact that Eq.~(\ref{eq:qm-as-id}) fails in lattice O6
(Fig.~\ref{fig:o6mo2}a), meaning it implies the orthomodular law by
Theorem 2 of \cite[p.~22]{kalmb83}.
\begin{definition}\label{def:ident}
\begin{eqnarray}
a\equiv b\ {\buildrel\rm def\over =}\ (a\cap b)\cup(a'\cap b')\label{eq:equiv}
\end{eqnarray}
\end{definition}
We note that $a\equiv b=a\equiv_5 b$ holds in all OMLs, so these
two identities
may be viewed as alternate definitions for the same operation in
OMLs.  The equality also holds in lattice O6, so they may
be used interchangeably in any orthomodular law equivalent
added to ortholattices; in particular
$\equiv$ may be substituted for $\equiv_5$ in the $i=5$ case of
Eq.~(\ref{eq:qm-as-id}). However $a\equiv b=a\equiv_5 b$ does
not hold in all ortholattices as shown in \cite{mpcommp99}, so
the two identities should be considered to be different operations from an
ortholattice point of view.

\begin{definition}\label{def:commut} We say that $a$ and $b$
{\em commute} in {\em OML} and write $aCb$ when either of the
following equations hold: \cite{holl95,zeman}
\begin{eqnarray}
a=((a\cap b)\cup (a\cap b'))\label{eq:commut1}\\
a\cap(a'\cup b)\le b\label{eq:commut2}
\end{eqnarray}
\end{definition}

\begin{lemma}\label{lemma:commut}
An {\em OL} in which Equations (\ref{eq:commut1}) and
(\ref{eq:commut2}) follow from each other is an OML.
\end{lemma}

Yet other forms of the orthomodularity condition are the following ones.

\begin{lemma}\label{lemma:oml-o} An {\em OL} in which any one of the following
conditions hold is an {\em OML} and vice versa.
\begin{eqnarray}
a\to_i b=1\qquad\Leftrightarrow\qquad a\le b,\qquad\qquad i=1,\dots,5
\label{eq:oml-le}
\end{eqnarray}
\end{lemma}
\begin{proof}The proof of Eq.~(\ref{eq:oml-le}) is given in \cite{pav87}
and \cite{pav89}. We stress that $\Leftarrow$ direction holds in any OL.
\end{proof}

Later on we shall make use of a definition based on the
following ``transitivity'' theorem which does not work for
$\equiv_i,\ i=1,\dots,4$. Note that in any OML
$a\equiv b=(a\cap b)\cup(a'\cap b')=(a\to_i b)\cap(b\to_i a)$ for
$i=1,\dots,5$ and that instead of the above Definition \ref{mixed-id}
one can use one with
$a\equiv_1 b=(a\to_1 b)\cap(b\to_4 a)$, etc. \cite{mpcommp99}

\medskip\noindent

\begin{theorem}\label{th:other-eq} An ortholattice in which
\begin{eqnarray}
(a\equiv  b)\cap (b\equiv  c)&\le&a\equiv  c\label{eq:om-alt}\\
(a\equiv  b)\cap (b\equiv  c)&=&(a\equiv  b)\cap (a\equiv  c)
\label{eq:om-alte}
\end{eqnarray}
hold is an orthomodular lattice and vice versa.

The same statement holds for $(a\to_i b)\cap(b\to_i a),\ i=1,\dots,5$
being substituted for $a\equiv b$.
\end{theorem}
\begin{proof}
Equations (\ref{eq:om-alt}) and (\ref{eq:om-alte}) fail in lattice O6,
so they imply the orthomodular law.

For the converse with Eq.~(\ref{eq:om-alt}) we start with
$((a\cap b)\cup(a'\cap b'))\cap(b'\cup(b\cap c))$. It is easy to show
that $(a'\cap b')C(b'\cup(b\cap c))$ and $(a'\cap b')C(a\cup b)$.
By applying the Foulis-Holland theorem (which we shall subsequently
refer to as F-H) \cite{holl95} to our
starting expression we obtain: $((a\cap b)\cap
(b'\cup(b\cap c)))\cup((a'\cap b')\cap
(b'\cup(b\cap c)))$. The first conjunction is by orthomodularity
equal to $a\cap b\cap c$. The disjunction is thus equal or less
than $a'\cup(a\cap c)$ and we arrive at
$(a\equiv b)\cap (b\to_1 c)\le (a\to_1
c)$. By multiplying both sides by $(c\to_1 b)$ we get
$(a\equiv b)\cap (b\equiv c)\le (c\to_1 b)\cap(a\to_1 c)
\le (a\to_1 c)$. By symmetry we also have $(a\equiv b)\cap
(b\equiv c)\le (c\to_1 a)$. A combination of the latter two
equations proves the theorem.  We draw the reader's attention
to the fact that $(a\to_1 b)\cap (b\equiv c)\le (a\to_1
c)$ does not hold in all OMLs (it is violated by MO2).
For the converse with Eq.~(\ref{eq:om-alte}) we start with
Eq.~(\ref{eq:om-alt}) and obtain $(a\equiv  b)\cap (b\equiv  c)
\le(a\equiv  b)\cap(a\equiv  c)$. On the other hand, starting
with $(a\equiv  b)\cap (a\equiv  c)\le(b\equiv  c)$
we obtain $(a\equiv  b)\cap(b\equiv  c)\le(a\equiv  b)
\cap(a\equiv  c)$. Therefore the conclusion.

As for the statements with $(a\to_i b)\cap(b\to_i a),\ i=1,\dots,5$
substituted for $a\equiv b$ they fail in O6, so they imply the
orthomodular law. For the converse it is sufficient to note that
in any OML the following holds: $(a\to_i b)\cap(b\to_i a)=
a\equiv b,\ i=1,\dots,5$
 \end{proof}

We conclude this section with an intriguing open problem whose
partial solutions we find with the help of states defined on
OML in the next section.
\begin{theorem}\label{th:ident-distr-eq}In any {\em OML} the following
conditions follow from each other.
\begin{eqnarray}
(a\equiv b)\cap((b\equiv c)\cup (a\equiv c))&\le&
(a\equiv  c)\label{eq:om-alt-v}\\
(a\equiv b)\to_0((a\equiv c)\equiv(b\equiv c))&=&1\label{eq:woml2}\\
(a\equiv b)\cap((b\equiv c)\cup(a\equiv c))&=&
((a\equiv b)\cap(b\equiv c))\cup
((a\equiv b)\cap(a\equiv c))\label{eq:id-distr}
\end{eqnarray}
Eq.~(\ref{eq:om-alt-v}) fails in {\em O6}. Eqs.~(\ref{eq:woml2}) and
(\ref{eq:id-distr}) fail in
lattices in which weakly {\em OML} ({\em WOML}) fail \cite{mphpa98}
but do not fail in {\em O6}.
\end{theorem}
\begin{proof}To obtain Eq.~(\ref{eq:woml2}) from Eq.~(\ref{eq:om-alt-v})
we apply Lemma \ref{lemma:oml-o} ($i=1$):
$1=(a\equiv b)'\cup((b\equiv c)'\cap(a\equiv c)')
\cup(((a\equiv c)\cap(a\equiv b))\cap((b\equiv c)\cup(a\equiv c)))=[{\rm
Eq}.\>(\ref{eq:om-alte})]=(a\equiv b)'\cup((a\equiv c)\cap(b\equiv c))
\cup((a\equiv c)'\cap(b\equiv c)')$.  Reversing the steps yields
(\ref{eq:om-alt-v}) from (\ref{eq:woml2}).

To get  Eq.~(\ref{eq:id-distr}) we first note that one can easily derive
$(a\equiv b)\cap((b\equiv c)\cup(a\equiv c))=(a\equiv c)\cap(a\equiv b)$
from Eq.~(\ref{eq:om-alt-v}). Then one gets (\ref{eq:id-distr}) by
applying Eq.~(\ref{eq:om-alte}).

To arrive at Eq.~(\ref{eq:om-alt-v}) starting from
Eq.~(\ref{eq:id-distr}) we apply Eq.~(\ref{eq:om-alte}) and reduce the
right hand side of Eq.~(\ref{eq:id-distr}) to
$(a\equiv b)\cap(a\equiv c)$ what yields Eq.~(\ref{eq:om-alt-v}).
\end{proof}

An open problem is whether conditions (\ref{eq:woml2}) and
(\ref{eq:id-distr}) hold in any WOML and whether these conditions
together with condition (\ref{eq:om-alt-v}) hold in any OML.
Note that Eq.~(\ref{eq:om-alt-v}) fails in O6 only because
Eqs.~(\ref{eq:om-alte}) and (\ref{eq:om-alt}), which we used to
infere it from Eq.~(\ref{eq:id-distr}), fail in O6.
We scanned all available orthomodular Greechie lattices\footnote{We
obtain the Greechie lattices with practically arbitrary number
of atoms and blocks by using the technique of \it isomorph-free
exhaustive generation\/\rm\cite{bdm-ndm-mp-1}. The reader can retrieve
many lattices with up to 38 atoms and blocks at
{\tt ftp://cs.anu.edu.au/pub/people/bdm/nauty/greechie.html}
and {\tt ftp://m3k.grad.hr/pavicic/greechie/diagrams} (legless),
and a program for making any desired set of lattices written in C by
B.~D.~McKay at {\tt ftp://m3k.grad.hr/pavicic/greechie/program}}
with up to 14 blocks (without legs and with 3 atoms in a block;
this makes 271930 legless lattices), over 400000 lattices with up to
17 blocks, and selected lattices with up to 38 blocks, but there was
no violation of any of them by the Eq.~(\ref{eq:om-alt-v}), so,
there is a strong indication that these conditions might hold in
any OML, but we were not able to either prove or disprove this.
We think it is an intriguing problem because repeated attempts to
prove these conditions either in WOML, or in OML, or in Hilbert space
always brought us into a kind of a vicious circle and also because we
were unable to prove that an even weaker condition holds in any OML.
We have, however, proved that the latter condition holds in Hilbert
space and we give the proof in the next section
(Equation \ref{eq:god-alt-v2}).

\section{\large States and Their Equations}
\label{sec:states}

In the standard approach of reconstructing Hilbert space one starts
from an orthomodular lattice, OML, then defines a state on OML,
and imposes additional conditions on the state as well on
OML to eventually arrive at the Hilbert space representation of
such a mixed lattice-state structure. [To get an insight into the
latter structure, below we first define a state on a lattice and
pinpoint a difference between classical and quantum strong states
(Definitions \ref{def:state} and \ref{def:state} and
Theorems \ref{th:strong-distr}, \ref{hilb-strong-s},
and \ref{th:god-eq}).]

Alternatively one can reconstruct Hilbert space solely by means of the
lattice theory. We start with an ortholattice, OL, build the Hilbert
lattice (Definition \ref{def:hl} and Theorem \ref{th:repr})
and with the help of three additional axioms arrive at its complex
Hilbert space representation (Theorem \ref{th:sol}) without invoking
the notion of state at all. The states needed for obtaining mean
values of measured observables follow from Gleason's theorem.

Going back to the traditional approach we explore how far one can
go in reconstructing the Hilbert space starting with a strong state
defined on OL without invoking any further either lattice or state
condition. We show that strong quantum states imposed on OL turn the
latter into an OML in which the so-called Godowski equations hold and
obtain several new traits of the equations and much simpler than
original Greechie lattices that characterize them (Theorems and
Lemmas \ref{th:god-eq}--\ref{th:go2n} and
\ref{th:new3go}--\ref{lem:goswap}. In the end we derive Mayet's
equations from Godowski's (Theorem \ref{th:mayet}).

\begin{definition}\label{def:state} A state on a lattice {\rm L}
 is a function $m:{\rm L}\longrightarrow [0,1]$
(for real interval $[0,1]$) such
that $m(1)=1$ and $a\perp b\ \Rightarrow\ m(a\cup b)=m(a)+m(b)$,
where $a\perp b$ means $a\le b'$.
\end{definition}

This implies $m(a)+m(a')=1$ and $a\le b\ \Rightarrow\ m(a)\le
m(b)$.

\begin{definition}\label{def:strong} A nonempty set $S$ of
states on {\rm L} is called a
strong set of {\em classical\/} states if
\begin{eqnarray}
(\exists m \in S)(\forall a,b\in{\rm L})((m(a)=1\ \Rightarrow
\ m(b)=1)\ \Rightarrow\ a\le b)\,\label{eq:st-cl}
\end{eqnarray}
and a strong set of {\em quantum\/} states if
\begin{eqnarray}
(\forall a,b\in{\rm L})(\exists m \in S)((m(a)=1\ \Rightarrow
\ m(b)=1)\ \Rightarrow\ a\le b)\,.\label{eq:st-qm}
\end{eqnarray}
We assume that {\rm L} contains more than one element and that
an empty set of states is not strong.  Whenever we omit the
word ``quantum'' we mean condition (\ref{eq:st-qm}).
\end{definition}

The first part of Definition \ref{def:strong} we have not seen in the
literature but consider it worth defining it because of the following
theorem.

\begin{theorem}\label{th:strong-distr} Any ortholattice that admits a
strong set of classical states is distributive.
\end{theorem}

\begin{proof} Eq.~(\ref{eq:st-qm}) follows from Eq.~(\ref{eq:st-cl})
and by Theorem \ref{th:god-eq} an ortholattice that admits a strong
set of classical states is orthomodular. Let now $a$ and $b$ be
any two lattice elements.  Assume, for state $m$,
that $m(b)=1$.  Since the lattice admits a strong set of classical
states, this implies $b=1$, so $m(a\cap b)=m(a\cap 1)=m(a)$.
But $m(a')+m(a)=1$ for any state, so $m(a\to_1 b)=m(a')+m(a\cap b)=1$.
Hence we have $m(b)=1\Rightarrow m(a\to_1 b)=1$, which means (since
the ortholattice admits a strong set of classical states)
that $b\le a\to_1b$.  This is another way of saying $aCb$. \cite{zeman}
By F-H, an orthomodular lattice in which any
two elements commute is distributive.
\end{proof}

We see that that a description of any classical measurement by
a classical logic (more precisely by its lattice model, a
Boolean algebra) and by a classical probability theory
coincide, because we can always find a single state (probability
measure) for all lattice elements. As opposed to this, a
description of any quantum measurement consists of two
inseparable parts: a quantum logic (i.e., its lattice model,
an orthomodular lattice) and a quantum probability theory,
because we must obtain different states for different lattice elements.

In order to enable an isomorphism between an orthocomplemented
orthomodular lattice and the corresponding Hilbert space we have
to add further conditions to the lattice. These conditions correspond
to the essential properties of any quantum system such as superposition
and make the so-called Hilbert lattice as
follows. \cite{holl95,ivertsj}

\begin{definition}\label{def:hl}An {\em OML} which satisfies the
following conditions is a Hilbert lattice, {\em HL}.
\begin{enumerate}
\item {\em Completeness:\/} The meet and join of any subset of
an {\em HL} always exist.
\item {\em Atomic:\/} Every non-zero element in an {\em HL} is greater
than or equal to an atom. (An atom $a$ is a non-zero lattice element
with $0< b\le a$ only if $b=a$.)
\item {\em Superposition Principle:\/} (The atom $c$ is a superposition
of the atoms $a$ and $b$ if $c\ne a$, $c\ne b$, and $c\le a\cup b$.)
\begin{enumerate}
\item Given two different atoms $a$ and $b$, there is at least
one other atom $c$, $c\ne a$ and $c\ne b$, that is a superposition
of $a$ and $b$.
\item If the atom $c$ is a superposition of  distinct atoms
$a$ and $b$, then atom $a$ is a superposition of atoms $b$ and $c$.
\end{enumerate}
\item {\em Minimal length:\/} The lattice contains at least
three elements $a,b,c$ satisfying: $0<a<b<c<1$.
\end{enumerate}
\end{definition}

Note that atoms correspond to pure states when defined on the
lattice. We recall that the {\it irreducibility\/}
and the {\it covering property\/} follow from the
superposition principle. \cite[pp.~166,167]{beltr-cass-book}
We also recall that any Hilbert lattice must contain a countably
infinite number of atoms. \cite{ivertsj} The above conditions
suffice to establish isomorphism between HL and the closed subspaces
of any Hilbert space, ${\cal C}({\cal H})$, through the following
well-known theorem. \cite[\S\S33,34]{maeda}

\begin{theorem}\label{th:repr}For every Hilbert lattice
{\rm HL} there exists a field $\cal K$ and a Hilbert space
$\cal H$ over $\cal K$, such that ${\cal C}({\cal H})$ is
ortho-isomorphic to {\rm HL}.

Conversely, let $\cal H$ be an infinite-dimensional Hilbert space
over a field $\cal K$ and let
\begin{eqnarray}
{\cal C}({\cal H})\ {\buildrel\rm def\over =}\ \{ {\cal X}\
\subset {\cal H}\ | \>{\cal X}^{\perp\perp}={\cal X}\}
\end{eqnarray}
be the set of all biorthogonal closed subspaces of $\cal H$.
Then ${\cal C}({\cal H})$ is a Hilbert lattice
relative to:
\begin{eqnarray}
a\cap b\ =\ {\cal X}_a\cap {\cal X}_b
\qquad\qquad {\rm and}\qquad\qquad a\cup b\
 =\ ({\cal X}_a+{\cal X}_b)^{\perp\perp}.
\end{eqnarray}
\end{theorem}

In order to determine the field over which Hilbert space
in Theorem \ref{th:repr} is defined we make use of the following
theorem.

\begin{theorem}\label{th:sol}{\em [Sol\`er-Mayet-Holland]}
Hilbert space $\cal H$ from Theorem \ref{th:repr} is
an infinite-dimensional one defined over a complex field
${\mathbb C}$ if the following conditions are met:

5. {\em Infinite orthogonality:} Any {\rm HL} contains a countably
infinite sequence of orthogonal elements.  \cite{soler}

6. {\em Unitary orthoautomorphism:} For any two orthogonal atoms
$a$ and $b$ there is an automorphism $\cal U$ such that ${\cal U}(a)
=b$, which satisfies ${\cal U}(a')={\cal U}(a)'$, i.e., it is an
{\em orthoautomorphism}, and whose mapping into $\cal H$ is a unitary
operator $U$ and therefore we also call it {\em unitary}.
\cite{holl95}

7. {\em ${\mathbb C}$ characterization:} There are pairwise
orthogonal elements $a,b,c\in {\rm L}$ such that
$(\exists d,e\in {\rm L})(0<d<a\ \&\ 0<e<b)$ and there is an
automorphism $\cal V$ in {\rm L} such that $({\cal V}(c)<c)$,
$(\forall f\in {\rm L}:f\le a)({\cal V}(f)=f)$,
$(\forall g\in {\rm L}:g\le b)({\cal V}(g)=g)$
and $(\exists h\in {\rm L})(
0\le h\le a\cup b\ \&\ {\cal V}({\cal V}(h))\ne h)$. \cite{mayet98}
\end{theorem}

\begin{proof} \cite{holl95} By Theorem \ref{th:repr}, to any two
orthogonal atoms $a$ and $b$ there correspond orthogonal one-dimensional
subspaces (vectors) $e$ and $f$ from $\cal H$ such that $a={\cal K}e$
and $b={\cal K}f$. The unitary orthoautomorphism $\cal U$ maps into
the unitary operator $U$ so as to give $U(e)=\alpha f$ for some
$\alpha\in\cal K$. {}From this and from the unitarity of $U$ we get:
$\langle e,e\rangle=\langle U(e),U(e)\rangle=
\langle\alpha f,\alpha f\rangle=\alpha\langle f,f\rangle\alpha^*$.
Hence, there is an infinite orthogonal sequence $\{e_i:i=1,2,\dots\}$,
such that $\langle e_i,e_i\rangle=\langle f_j,f_j\rangle$, for all
$i,j$. Then Sol\`er's \cite{soler} and Mayet's   \cite{mayet98}
theorems prove the claim.
\end{proof}

We have seen that the definition of the ``unitarity'' of the
{\em unitary automorphism} in the previous theorem is not given
directly in HL but through the inner product of the corresponding
Hilbert space (whose existence is guaranteed by Theorem
\ref{th:repr}). A pure lattice version of the definition of
the {\em unitary automorphism} formulated by S.~S.~Holland, JR.,
\cite{holl95} is not known, but it is known that it can be
replaced by Morash's purely lattice theoretic {\em angle bisecting}
condition in HL. \cite{soler}

{}From the previous two theorems we see that to arrive at the basic
Hilbert space structure we do not need the notion of state, i.e.,
of the probability of geting a value of a measured observable.
This probability and state follow uniquely from Hilbert space by
Gleason's theorem and we can use them to make probabilistic (the only
available ones in the quantum theory) predictions of an observable
$\cal A$: $Prob({\cal A})=tr({\mathbf\rho}{\cal A})$, there $tr$ is
the trace and $\mathbf\rho$ is a density matrix. \cite[p.~178]{isham}
Alternatively we can start with the pure states that correspond to one
dimensional subspaces of Hilbert space, i.e., to vectors of Hilbert
space and to  atoms in the Hilbert lattice.

\begin{definition}\label{def:pure-st} A state $m$ is called {\em pure} if,
for all states $m_1,m_2$ and all reals $0<\lambda<1$, the equality
$m=\lambda m_1+(1-\lambda)m_2$ implies $m=m_1=m_2$
\end{definition}

According to Gleason's theorem \cite{gleason}, for
every vector $\Psi_m\in{\cal H},\ \|\Psi_m\|=1$ and for every
$P_a$, where $P_a$ is a projector on the subspace ${\cal H}_a$,
there exists a unique inner product $\langle P_a\Psi_m,\Psi_m\rangle$
which is a pure state $m(a)$ on ${\cal C}({\cal H})$.
By the spectral theorem to each subspace there
corresponds a self-adjoint operator $\cal A$ and we write
$P_a=P_{\cal A}$. The mean value of $\cal A$ in the state $m$ is
$\langle{\cal A}\rangle={\rm Exp}\,m({\cal A})=\int\,\alpha d \langle
P_{{\cal A},\{\alpha\}}\Psi_m,\Psi_m\rangle=\langle
{\cal A}\Psi_m,\Psi_m\rangle$.
\cite{maczin}

So, conditions 1-4 of Definition \ref{def:hl} and 5-7 from
Definition \ref{th:sol} enable a one-to-one correspondence between
the lattice elements and the closed subspaces of the infinite-dimensional
Hilbert space of a quantum system and Gleason's theorem enables
a one-to-one correspondence between states and mean
values of the operators measured on the system provided the above
strong states (probability measures) are defined on them.
The usage of strong states here is somewhat unusual because most
authors use full states instead. \cite{holl95,maczin,kalmb83}
To prove that the correspondence (isomorphism) holds for the strong
states as well, we only have to prove that Hilbert space admits
strong states because the other direction follows from the fact that
any strong set of states is full. The result is not new (it
appears, e.g., in \cite[p.~144]{beltr-cass-book}) but we give here a
proof communicated to us by Ren\'e Mayet, for the sake of completeness.

\begin{theorem}\label{hilb-strong-s} Any Hilbert lattice admits a
strong set of states.
\end{theorem}
\begin{proof} We need only to use pure states defined by unit vectors:
If $a$ and $b$ are closed subspaces of Hilbert space, $\cal H$
such that $a$ is not contained in $b$, there is a unit vector $u$ of
$\cal H$ belonging to $a-b$. If for each $c$ in the lattice of all
closed subspaces of $\cal H$, ${\cal C}({\cal H})$,
we define $m(c)$ as the square of the norm of
the projection of $u$ onto $c$, then $m$ is a state on $\cal H$ such
that $m(a)=1$ and $m(b)<1$. This proves that ${\cal C}({\cal H})$
admits a strong set of states, and this proof works in each of the
3 cases where the underlying field is the field of real numbers,
of complex numbers, or of quaternions.

We can formalize the proof as follows:
\begin{eqnarray}
&&(\forall a,b\in L)((\sim\ a\le b)\ \Rightarrow\
(\exists m\in S)(m(a)=1\ \&\ \sim\ m(b)=1))\nonumber\\
&\Rightarrow & (\forall a,b\in L)(\exists m\in S)((m(a)=1\
\Rightarrow\ m(b)=1)\ \Rightarrow\ a\le b)\nonumber
\end{eqnarray}
\end{proof}

So, any Hilbert space admits strong states and we need them to predict
outcomes of measurements. But there is more to it---states, when
defined on an ortholattice, impose very strong conditions on it.
In particular, they impose a class of orthomodular
equations which hold in ${\cal C}({\cal H})$ and do not hold in all
OMLs: Godowski's \cite{godow} and Mayet's \cite{mayet86} equations.
In the rest of this section we shall first give some alternative
formulations of Godowski's equations and present a new class of
lattices in which the equations fail. Then we shall show that Mayet's
Examples 2, 3, and 4 which were meant to illustrate a generalization
of Godowski's equations are nothing but special cases of the latter
equations.

\begin{definition}\label{def:god-equiv}Let us call the
following expression the {\em Godowski identity}:
\begin{eqnarray}
a_1{\buildrel\gamma\over\equiv}a_n{\buildrel{\rm def}
\over =}(a_1\to_1a_2)\cap(a_2\to_1a_3)\cdots
\cap(a_{n-1}\to_1a_n)\cap(a_n\to_1a_1),\qquad
n=3,4,5,\dots\label{eq:god-equiv}
\end{eqnarray}
\end{definition}
We define $a_n{\buildrel\gamma\over\equiv}a_1$ in the same way with
variables $a_i$ and $a_{n-i+1}$ swapped;
in general $a_i{\buildrel\gamma\over\equiv}a_j$ will be an expression
with $|j-i|+1\ge 3$ variables $a_i,\ldots,a_j$ first appearing in that
order.  For completeness and later use (Theorem \ref{th:god-trans}) we
define $a_i{\buildrel\gamma\over\equiv}a_i{\buildrel{\rm def} \over
=} (a_i\to_1 a_i)=1$ and
$a_i{\buildrel\gamma\over\equiv}a_{i+1}{\buildrel{\rm def} \over
=}(a_i\to_1 a_{i+1})\cap(a_{i+1}\to_1 a_i)=a_i\equiv a_{i+1}$, the last
equality holding in any OML.  We also define
$a_1{\buildrel\delta\over\equiv}a_n$, etc.\ with the substitution of
$\to_2$ for $\to_1$ in $a_1{\buildrel\gamma\over\equiv}a_n$, etc.

\begin{theorem}\label{th:god-eq} Godowski's equations {\em\cite{godow}}
\begin{eqnarray}
a_1{\buildrel\gamma\over\equiv}a_3
&=&a_3{\buildrel\gamma\over\equiv}a_1
\label{eq:godow3o}\\
a_1{\buildrel\gamma\over\equiv}a_4
&=&a_4{\buildrel\gamma\over\equiv}a_1
\label{eq:godow4o}\\
a_1{\buildrel\gamma\over\equiv}a_5
&=&a_5{\buildrel\gamma\over\equiv}a_1
\label{eq:godow5o}\\
&\dots &\nonumber
\end{eqnarray}
hold in all ortholattices, {\em OL}'s with strong sets of states.
An {\em OL} to which these equations are added is a variety
smaller than {\em OML}.

We shall call these equations {\em $n$-Go} ({\em 3-Go}, {\em 4-Go},
etc.).  We also denote by {\em $n$GO} ({\em 3GO}, {\em 4GO}, etc.) the
{\em OL} variety determined by {\em $n$-Go} (which we also call the {\em
$n$GO law}).
\end{theorem}
\begin{proof} The proof is similar to that in \cite{godow}. By
Definition \ref{def:state} we have
$m(a_1\to_1a_2)=m(a'_1)+m(a_1\cap a_2)$ etc.,
because $a'_1\le(a'_1\cup a'_2)$, i.e., $a'_1\perp(a_1\cap a_2)$
in any ortholattice. Assuming $m(a_1{\buildrel\gamma\over\equiv}a_n)=1$
we get $m(a_1\to_1a_2)=\cdots =m(a_{n-1}\to_1a_n)=m(a_n\to_1a_1)=1$.
Hence, $n=m(a_1\to_1a_2)\cdots+m(a_{n-1}\to_1a_n)+m(a_n\to_1a_1)=
m(a_n\to_1a_{n-1})\cdots+m(a_2\to_1a_1)+m(a_1\to_1a_n)$. Therefore,
$m(a_n\to_1a_{n-1})=\cdots=m(a_2\to_1a_1)=m(a_1\to_1a_n)=1$. Thus,
by Definition \ref{def:strong} for strong quantum states, we obtain:
$(a_1{\buildrel\gamma\over\equiv}a_n)\le (a_n\to_1a_{n-1})$,\ \ldots,
$(a_1{\buildrel\gamma\over\equiv}a_n)\le (a_2\to_1a_1)$, and
$(a_1{\buildrel\gamma\over\equiv}a_n)\le (a_1\to_1a_n)$, wherefrom
we get $(a_1{\buildrel\gamma\over\equiv}a_n)\le
(a_n{\buildrel\gamma\over\equiv}a_1)$. By symmetry, we get
$(a_n{\buildrel\gamma\over\equiv}a_1)\le
(a_1{\buildrel\gamma\over\equiv}a_n)$. Thus
$(a_1{\buildrel\gamma\over\equiv}a_n)=
(a_n{\buildrel\gamma\over\equiv}a_1)$.

$n$GO is orthomodular because 3-Go fails in O6, and $n$-Go implies
$(n-1)$-Go in any OL (Lemma \ref{lem:god-iimpliesn-1}). It is a variety
smaller than OML because 3-Go fails in the Greechie lattice from
Fig.~\ref{fig-oag34}a.
\end{proof}

The following lemma provides a result we will need.
\begin{lemma}\label{lem:god-prelemma3a}
The following equation holds in all {\em OML}s.
\begin{eqnarray}
&(a_1\equiv a_2)\cdots\cap(a_{n-1}\equiv a_n)
=(a_1\cdots\cap a_n)\cup(a_1'\cdots\cap a_n'),\quad n\ge 2
\label{eq:god-prelemma3a}
\end{eqnarray}
\end{lemma}
\begin{proof}
We use induction on $n$.  The basis is
simply the definition of $\equiv$.  Suppose
$(a_1\equiv a_2)\cdots\cap(a_{n-2}\equiv a_{n-1})
=(a_1\cdots\cap a_{n-1})\cup(a_1'\cdots\cap a_{n-1}')$.
Multiplying both sides by $a_{n-1}\equiv a_n=
(a_{n-1}\to_1 a_n)\cap(a_n\to_2 a_{n-1})$, we have
\begin{eqnarray}
\lefteqn{
(a_1\equiv a_2)\cdots\cap(a_{n-1}\equiv a_n)}\nonumber\\
&&\qquad=[((a_1\cdots\cap a_{n-1})\cup(a_1'\cdots\cap a_{n-1}'))\cap
(a_{n-1}\to_1 a_n)]\cap(a_n\to_2 a_{n-1})\nonumber\\
&&\qquad=[(a_1\cdots\cap a_n)\cup(a_1'\cdots\cap a_{n-1}')
]\cap(a_n\to_2 a_{n-1})\nonumber\\
&&\qquad=(a_1\cdots\cap a_n)\cup(a_1'\cdots\cap a_n')\,.\nonumber
\end{eqnarray}
F-H was used in the last
two steps, whose details we leave to the reader.
\end{proof}

\begin{theorem}\label{th:god-th1} An {\em OL} in which any of
the following equations holds is an {\em $n$GO} and vice versa.
\begin{eqnarray}
a_1{\buildrel\delta\over\equiv}a_n
&=&a_n{\buildrel\delta\over\equiv}a_1
\label{eq:godow2}\\
a_1{\buildrel\gamma\over\equiv}a_n
&=&(a_1\equiv a_2)\cap(a_2\equiv a_3)\cdots\cap(a_{n-1}\equiv a_n)
\label{eq:godow1c}\\
a_1{\buildrel\delta\over\equiv}a_n
&=&(a_1\equiv a_2)\cap(a_2\equiv a_3)\cdots\cap(a_{n-1}\equiv a_n)
\label{eq:godow2c}\\
a_1{\buildrel\gamma\over\equiv}a_n
&\le&a_1\to_ia_n,\qquad i=1,2,3,5
\label{eq:godow1d}\\
a_1{\buildrel\delta\over\equiv}a_n
&\le&a_1\to_ia_n,\qquad i=1,2,4,5
\label{eq:godow2d}\\
(a_1{\buildrel\gamma\over\equiv}a_n)\cap(a_1\cup a_2\cdots\cup a_n)
&=&a_1\cap a_2\cdots\cap a_n
\label{eq:godow1e}\\
(a_1{\buildrel\delta\over\equiv}a_n)\cap(a_1'\cup a_2'\cdots\cup a_n')
&=&a_1'\cap a_2'\cdots\cap a_n'
\label{eq:godow2e}
\end{eqnarray}
\end{theorem}
\begin{proof}
Lattice O6 violates all of the above equations as well as $n$-Go.  Thus
for the proof we can presuppose that any OL in which they hold is an
OML.

Equation (\ref{eq:godow2}) follows from definitions, replacing
variables with their orthocomplements in $n$-Go.

Assuming (\ref{eq:godow1c}), we make use of
$a\equiv b= (a\to_1 b)\cap(b\to_1 a)$
to obtain the equivalent equation $a_1{\buildrel\gamma\over\equiv}a_n=
a_1{\buildrel\gamma\over\equiv}a_n\cap
a_n{\buildrel\gamma\over\equiv}a_1$, so
$a_1{\buildrel\gamma\over\equiv}a_n\le
a_n{\buildrel\gamma\over\equiv}a_1$.  By renaming
variables, the other
direction of the inequality also holds, establishing $n$-Go.
Conversely, $n$-Go immediately implies
$a_1{\buildrel\gamma\over\equiv}a_n=
a_1{\buildrel\gamma\over\equiv}a_n\cap
a_n{\buildrel\gamma\over\equiv}a_1$.  The proof
for (\ref{eq:godow2c}) is similar.

For (\ref{eq:godow1d}) and (\ref{eq:godow2d}), we demonstrate only
(\ref{eq:godow1d}), $i=3$.  {}From (\ref{eq:godow1d}), by rearranging
factors on the left-hand-side we have
$a_1{\buildrel\gamma\over\equiv}a_n \le
a_2\to_3a_1$, so
$a_1{\buildrel\gamma\over\equiv}a_n\le (a_2\to_3a_1)
\cap (a_1\to_1a_2) = a_1\equiv a_2$ (from Table 1 in
\cite{mpcommp99}), etc.; this way we build up (\ref{eq:godow1c}).  For
the converse, (\ref{eq:godow1d}) and (\ref{eq:godow2d}) obviously follow
from (\ref{eq:godow1c}) and (\ref{eq:godow2c}).

For (\ref{eq:godow1e}), using (\ref{eq:god-prelemma3a}) we can write
(\ref{eq:godow1c}) as $a_1{\buildrel\gamma\over\equiv}a_n
=(a_1\cdots\cap a_n)\cup(a_1'\cdots\cap a_n')$.  Multiplying both sides
by $a_1\cdots\cup a_n$ and using F-H we obtain
$(\ref{eq:godow1e})$.  Conversely, disjoining both sides of
$(\ref{eq:godow1e})$ with $a_1'\cdots\cap a_n'$ and using
F-H and (\ref{eq:god-prelemma3a}) we obtain
(\ref{eq:godow1c}).  The proof for (\ref{eq:godow2e}) is similar.
\end{proof}

\begin{theorem}\label{th:god-th2}
In any {\rm $n$GO}, $n=3,4,5,\ldots$, all of the following equations hold.
\begin{eqnarray}
a_1{\buildrel\gamma\over\equiv}a_n
&\le&a_j\to_ia_k, \qquad 0\le i\le 5,\ 1\le j\le n,\ 1\le k\le n
\label{eq:godow1d2}\\
a_1{\buildrel\delta\over\equiv}a_n
&\le&a_j\to_ia_k, \qquad 0\le i\le 5,\ 1\le j\le n,\ 1\le k\le n
\label{eq:godow2d2}\\
a_1{\buildrel\gamma\over\equiv}a_n
&=&a_1{\buildrel\delta\over\equiv}a_n
\label{eq:godowf}
\end{eqnarray}
\end{theorem}
\begin{proof}
These obviously follow
from (\ref{eq:godow1c}) and (\ref{eq:godow2c}) and (for $i=0$)
the fact that $a\to_m b\le a\to_0 b,\ 0\le m\le 5$.
\end{proof}

Some of the equations of Theorem~\ref{th:god-th2} (in addition to
those mentioned in Theorem~\ref{th:god-th1}) also imply the $n$GO
laws.  In Theorem~\ref{th:god-theorem3} below we show them for $n=3$.
First we prove the following preliminary results.

\begin{lemma}\label{lem:god-prelemma3}
The following equations hold in all {\em OML}s.
\begin{eqnarray}
&(a\to_2 b)\cap(b\to_1 c)
=(a'\cap b')\cup(b\cap c)
\label{eq:god-prelemma3b}\\
&(a_1 \to_5 a_2)\cap(a_2 \to_5 a_3)\cap(a_3 \to_5 a_1)
=(a_1 \equiv a_2)\cap(a_2 \equiv a_3)
\label{eq:god-prelemma3c}
\end{eqnarray}
\end{lemma}
\begin{proof}
For (\ref{eq:god-prelemma3b}), $(a\to_2 b)\cap(b\to_1 c)=
((b\cup(a'\cap b'))\cap(b'\cup(b\cap c))=
((b\cup(a'\cap b'))\cap b')\cup((b\cup(a'\cap b'))\cap b\cap c)=
(a'\cap b')\cup(b\cap c)$.

For (\ref{eq:god-prelemma3c}), we have
\begin{eqnarray}
\lefteqn{(a_1 \to_5 a_2)\cap(a_2 \to_5 a_3)\cap(a_3 \to_5 a_1)}\nonumber\\
&&\qquad=[(a_1\equiv a_2)\cup(a_1'\cap a_2)]\cap[(a_2\to_1 a_3)
    \cap(a_2\to_2 a_3)]\cap[(a_3\to_1 a_1)\cap(a_3\to_2 a_1)]\nonumber\\
&&\qquad\le((a_1\equiv a_2)\cup(a_1'\cap a_2))\cap
    (a_2\to_2 a_3)\cap(a_3\to_1 a_1)\nonumber\\
&&\qquad=((a_1\equiv a_2)\cup(a_1'\cap a_2))
    \cap((a_2'\cap a_3')\cup(a_3\cap a_1))\nonumber\\
&&\qquad=((a_1\equiv a_2)\cap((a_2'\cap a_3')\cup(a_3\cap a_1)))\cup
    ((a_1'\cap a_2)\cap((a_2'\cap a_3')\cup(a_3\cap a_1)))\nonumber\\
&&\qquad=((a_1\equiv a_2)\cap((a_2'\cap a_3')\cup(a_3\cap a_1)))\cup
    0\nonumber\\
&&\qquad\le a_1\equiv a_2\,.\nonumber
\end{eqnarray}
In the third step we used (\ref{eq:god-prelemma3b}); in the fourth
$a_1\equiv  a_2 C a_1'\cap a_2$ and
$a_1'\cap a_2 C (a_2'\cap a_3')\cup(a_3\cap a_1)$; in the fifth
$a_1'\cap a_2 C a_2'\cap a_3'$ and
$a_1'\cap a_2 C a_3\cap a_1$.  Rearranging the left-hand side, this
proof also gives us $(a_1 \to_5 a_2)\cap(a_2 \to_5 a_3)\cap(a_3 \to_5 a_1)
\le a_2\equiv a_3$ and thus $\le (a_1 \equiv a_2)\cap(a_2 \equiv a_3)$.
The other direction of the inequality follows from $a\equiv b\le a\to_5 b$.
\end{proof}

\begin{theorem}\label{th:god-theorem3}
When $n=3$, an {\em OML} in which any of
the following equations holds is a {\em $n$GO} and vice versa.
\begin{eqnarray}
a_1{\buildrel\gamma\over\equiv}a_n
&\le&a_n\to_ia_1,\qquad i=2,3,4,5
\label{eq:godow3a}\\
a_1{\buildrel\delta\over\equiv}a_n
&\le&a_n\to_ia_1,\qquad i=1,3,4,5
\label{eq:godow3b}\\
a_1{\buildrel\gamma\over\equiv}a_n
&=&a_1{\buildrel\delta\over\equiv}a_n
\label{eq:godow3c}
\end{eqnarray}
\end{theorem}
\begin{proof}
We have already proved the converses in Theorem~\ref{th:god-th2}.

{}From (\ref{eq:godow3a}), we have $a_1{\buildrel\gamma\over\equiv}a_3
\le (a_3\to_i a_1)\cap(a_3\to_1 a_1)=a_3\to_5 a_1$ since
$(a\to_j b)\cap(a\to_k b)=a\to_5 b$ when $j\ne k$ for $j,k=1,\ldots,5$.
By rearranging the left-hand side we also have
$a_1{\buildrel\gamma\over\equiv}a_3\le a_1\to_5 a_2$ and
$\le a_2\to_5 a_3$.  Thus
$a_1{\buildrel\gamma\over\equiv}a_3\le
(a_1\to_5 a_2)\cap(a_2\to_5 a_3)\cap(a_3\to_5 a_1)\le a_1\equiv a_2
\le a_2\to_1 a_1$, which is the 3GO law by (\ref{eq:godow1d}).  In the
penultimate step we used (\ref{eq:god-prelemma3c}).

The proof for (\ref{eq:godow3b}) is similar, and from (\ref{eq:godow3c})
we obtain (\ref{eq:godow3a}).
\end{proof}
Whether Theorem \ref{th:god-theorem3} holds for $n>3$ is not known.

The equations obtained by substituting $\to_2$ for one or more $\to_1$'s
in Godowski's equations also hold in some
$n$GO, although to show such an equation with $j$ variables
may require the use of an $n$-Go equation with $n>j$.

\begin{theorem}\label{th:god-theorem4}
The following equation with $i$ variables holds in some {\em $n$GO}
with $n\ge i$, where
each $\to_{j_k}$ $(1\le k\le i)$ is either $\to_1$ or $\to_2$
in any combination.
\begin{eqnarray}
(a_1\to_{j_1}a_2)\cap(a_2\to_{j_2}a_3)\cdots
\cap(a_{i-1}\to_{j_{i-1}}a_i)\cap(a_i\to_{j_i}a_1)&=&
a_1{\buildrel\gamma\over\equiv}a_i
\label{eq:god-th4}
\end{eqnarray}
\end{theorem}
\begin{proof}
We illustrate the proof by showing that the 3-variable equation
\begin{eqnarray}
(a_1\to_2 a_2)\cap(a_2\to_1 a_3)\cap(a_3\to_1 a_1)&=&
a_1{\buildrel\gamma\over\equiv}a_3
\label{eq:th4-3}
\end{eqnarray}
holds in any 4GO.
The essential identities we
use are
\begin{eqnarray}
(a\to_1 (a\cup b))\cap((a\cup b)\to_1 b)&=&a\to_2 b\label{eq:th4a}\\
(a\to_2 (a\cap b))\cap((a\cap b)\to_2 b)&=&a\to_1 b\label{eq:th4b}
\end{eqnarray}
which hold in any OML.  Starting with (\ref{eq:godow1d2}), we have
\begin{eqnarray}
\lefteqn{
(a_1\to_1 b)\cap(b\to_1 a_2)\cap(a_2\to_1 a_3)\cap
(a_3\to_1 a_1)}\nonumber\\
&&\qquad\le (a_1\to_1 a_3)\cap(a_3\to_1 a_1)
  \cap(a_3\to_1 a_2)\cap(a_2\to_1 a_3)
\nonumber\\
&&\qquad=(a_1\equiv a_3)\cap(a_3\equiv a_2)
\nonumber\\
&&\qquad=(a_1\equiv a_2)\cap(a_2\equiv a_3)
\nonumber\\
&&\qquad=a_1{\buildrel\gamma\over\equiv}a_3\nonumber
\end{eqnarray}
where in the penultimate step we used (\ref{eq:om-alte})
[or more generally (\ref{eq:god-prelemma3a})] and in the last
step (\ref{eq:godow1c}). Substituting $a_1\cup a_2$ for $b$ and using
(\ref{eq:th4a}) we obtain
\begin{eqnarray}
(a_1\to_2 a_2)\cap(a_2\to_1 a_3)\cap(a_3\to_1 a_1)&\le&
a_1{\buildrel\gamma\over\equiv}a_3\nonumber
\end{eqnarray}
Using (\ref{eq:godow1d2}) for the other direction of the inequality,
we obtain (\ref{eq:th4-3}).
The reader should be able
to construct the general proof.
\end{proof}

A consequence of (\ref{eq:god-th4}) that holds in any 4GO is
\begin{eqnarray}
(a\to_1 b)\cap(b\to_2 c)\cap(c\to_1 a)&\le &(a\equiv c)
\label{eq:god-alt-v1}
\end{eqnarray}
which, using (\ref{eq:god-prelemma3b}) and weakening the left-most factor,
implies
\begin{eqnarray}
(a\equiv b)\cap((b'\cap c')\cup(a\cap c))&\le& (a\equiv c)\,.
\label{eq:god-alt-v2}
\end{eqnarray}
Equation (\ref{eq:god-alt-v2}) is also a consequence of
(\ref{eq:om-alt-v}) as can be seen if we write (\ref{eq:om-alt-v})
as follows:
\begin{eqnarray}
(a\equiv b)\cap((b\cap c)\cup(b'\cap c')\cup(a\cap c)\cup(a'\cap c'))
&\le& (a\equiv c)\,.\nonumber
\end{eqnarray}
As with (\ref{eq:om-alt-v}), we were unable to prove that even the
weaker-looking (\ref{eq:god-alt-v2}) holds in all OMLs.  It is also
unknown if (\ref{eq:god-alt-v2}) even holds in all 3GOs.  Finally, we do
not know if there is an $n$ such that (\ref{eq:om-alt-v}) holds in all
$n$GOs.

The equations obtained by substituting $\to_i$ for $\to_1$ in the
Godowski equations do not in general
result in equivalents for $i=3,4,5$ nor even hold in an $n$GO.
For example, for
$n=3$, the equations
$(a_1 \to_3 a_2)\cap(a_2 \to_3 a_3)\cap(a_3 \to_3 a_1)\le(a_2 \to_3 a_1)$
and
$(a_1 \to_4 a_2)\cap(a_2 \to_4 a_3)\cap(a_3 \to_4 a_1)\le(a_2 \to_4 a_1)$
fail in lattice MO2 ({\it Chinese
lantern}, Fig.~\ref{fig:o6mo2}b), and
$(a_1 \to_5 a_2)\cap(a_2 \to_5 a_3)\cap(a_3 \to_5 a_1)\le(a_2 \to_5 a_1)$
holds in all OMLs by (\ref{eq:god-prelemma3c}).

\begin{lemma}\label{lem:god-iimpliesn-1}
Any {\em $n$GO} is an {\em $(n-1)$GO}, $n=4,5,6,\ldots$.
\end{lemma}
\begin{proof}
Substitute $a_1$ for $a_2$ in equation $n$-Go.
\end{proof}

\begin{figure}[htbp]\centering
  \setlength{\unitlength}{1pt}
  \begin{picture}(240,120)(0,0)

    \put(5,13) { 
      \begin{picture}(124,110)(0,0) 
        \put(32.2,0){\line(1,0){55.6}}
        \put(32.2,100){\line(1,0){55.6}}
        \put(2.3,50){\line(3,5){29.9}}
        \put(117.7,50){\line(-3,5){29.9}}
        \put(2.3,50){\line(3,-5){29.9}}
        \put(117.7,50){\line(-3,-5){29.9}}
        \put(60,100){\line(0,-1){50}}
        \put(17.25,25){\line(5,3){42.75}}
        \put(102.75,25){\line(-5,3){42.75}}

        \put(2.3,50){\circle{10}}
        \put(117.7,50){\circle{10}}
        \put(87.8,0){\circle{10}}
        \put(87.8,100){\circle{10}}
        \put(32.2,0){\circle{10}}
        \put(32.2,100){\circle{10}}
        \put(60,0){\circle{10}}
        \put(60,100){\circle{10}}
        \put(17.25,25){\circle{10}}
        \put(17.25,75){\circle{10}}
        \put(102.75,25){\circle{10}}
        \put(102.75,75){\circle{10}}
        \put(60,50){\circle{10}}
        \put(38.9,37.5){\circle{10}}
        \put(81.4,37.5){\circle{10}}
        \put(60,75){\circle{10}}
      \end{picture}
    } 

    \put(170,0) { 
      \begin{picture}(124,110)(0,0) 
        \put(35.15,0){\line(1,0){49.7}}
        \put(35.15,120){\line(1,0){49.7}}
        \put(0,35.15){\line(0,1){49.7}}
        \put(120,35.15){\line(0,1){49.7}}
        \put(0,35.15){\line(1,-1){35.15}}
        \put(0,84.85){\line(1,1){35.15}}
        \put(120,35.15){\line(-1,-1){35.15}}
        \put(120,84.85){\line(-1,1){35.15}}
        \put(60,0){\line(1,6){12}}
        \put(0,60){\line(6,1){72}}
        \put(60,120){\line(1,-4){12}}
        \put(120,60){\line(-4,1){48}}

        \put(35.15,0){\circle{10}}
        \put(35.15,120){\circle{10}}
        \put(84.85,0){\circle{10}}
        \put(84.85,120){\circle{10}}
        \put(0,35.15){\circle{10}}
        \put(0,84.85){\circle{10}}
        \put(120,35.15){\circle{10}}
        \put(120,84.85){\circle{10}}
        \put(60,0){\circle{10}}
        \put(60,120){\circle{10}}
        \put(0,60){\circle{10}}
        \put(120,60){\circle{10}}
        \put(17.575,17.575){\circle{10}}
        \put(17.575,102.425){\circle{10}}
        \put(102.425,17.575){\circle{10}}
        \put(102.425,102.425){\circle{10}}
        \put(72,72){\circle{10}}
        \put(36,66){\circle{10}}
        \put(66,36){\circle{10}}
        \put(96,66){\circle{10}}
        \put(66,96){\circle{10}}

      \end{picture}
    } 

  \end{picture}
  \caption{\hbox to3mm{\hfill}(a)
   Greechie diagram for OML G3;
  \hbox to2mm{\hfill} (b) Greechie diagram for OML G4.
\label{fig-oag34}}
\end{figure}

The converse of Lemma (\ref{lem:god-iimpliesn-1}) does not hold.
Indeed, the {\it wagon wheel} OMLs G$n$, $n=3,4,5,...$, are related to
the $n$-Go equations in the sense that G$n$ violates $n$-Go but (for
$n\ge 4$) not $(n-1)$-Go.  In Fig.~\ref{fig-oag34} we show examples G3
and G4; for larger $n$ we construct G$n$ by adding more ``spokes'' in
the obvious way (according to the general scheme described in
\cite{godow}).

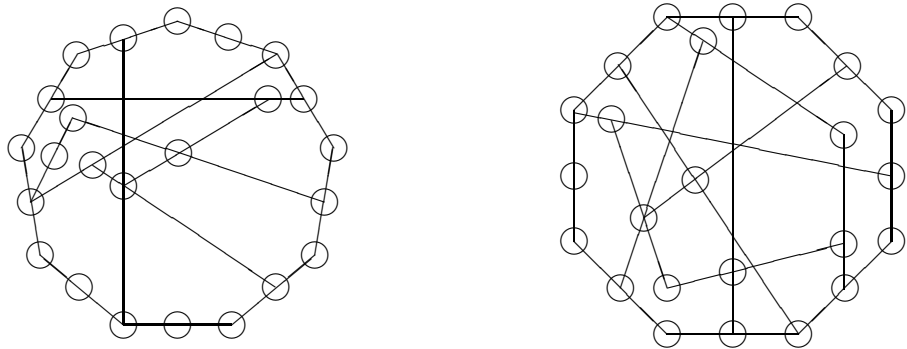
\begin{figure}[htbp]\centering
  \setlength{\unitlength}{1pt}
  \begin{picture}(240,120)(0,0)
    \put(-40,0) { 
      \begin{picture}(120,110)(0,0) 
        \put(39.48,3.62){\line(1,0){41.04}}
        \put(39.48,3.62){\line(-6,5){31.44}}
        \put(80.52,3.62){\line(6,5){31.44}}
        \put(8.04,30){\line(-1,6){6.75}}
        \put(111.96,30){\line(1,6){6.75}}
        \put(0.91,70.42){\line(3,5){21.2}}
        \put(119.09,70.42){\line(-3,5){21.2}}
        \put(22.13,105.96){\line(3,1){38}}
        \put(97.87,105.96){\line(-3,1){38}}
        \put(12.17,89.19){\line(1,0){95.3}}
        \put(4.475,50.21){\line(5,3){93.1}}
        \put(4.475,50.21){\line(1,2){15.8}}
        \put(115.525,50.21){\line(-3,1){95}}
        \put(28,64){\line(3,-2){69}}
        \put(39.48,56){\line(5,3){55.3}}
        \put(39.48,3.62){\line(0,1){108}}
        \put(80.52,3.62){\circle{10}}
        \put(39.48,3.62){\circle{10}}
        \put(111.96,30){\circle{10}}
        \put(8.04,30){\circle{10}}
        \put(119.09,70.42){\circle{10}}
        \put(0.91,70.42){\circle{10}}
        \put(97.2,105.8){\circle{10}}
        \put(21.8,105.8){\circle{10}}
        \put(60,118.6){\circle{10}}
        \put(60,3.62){\circle{10}}
        \put(97.24,17.81){\circle{10}}
        \put(22.76,17.81){\circle{10}}
        \put(115.525,50.21){\circle{10}}
        \put(4.475,50.21){\circle{10}}
        \put(107.83,89.19){\circle{10}}
        \put(12.17,89.19){\circle{10}}
        \put(79.285,112.5){\circle{10}}
        \put(39.48,112.3){\circle{10}}
        \put(94.3,89.19){\circle{10}}
        \put(60.4,68.5){\circle{10}}
        \put(39.48,56){\circle{10}}
        \put(28,64){\circle{10}}
        \put(20.5,82){\circle{10}}
        \put(13.4,67.6){\circle{10}}
      \end{picture}
    } 

    \put(170,0) { 
      \begin{picture}(120,110)(0,0) 
        \put(35.15,0){\line(1,0){49.7}}
        \put(35.15,120){\line(1,0){49.7}}
        \put(0,35.15){\line(0,1){49.7}}
        \put(120,35.15){\line(0,1){49.7}}
        \put(0,35.15){\line(1,-1){35.15}}
        \put(0,84.85){\line(1,1){35.15}}
        \put(120,35.15){\line(-1,-1){35.15}}
        \put(120,84.85){\line(-1,1){35.15}}
        \put(60,120){\line(0,-1){120}}
        \put(102,17){\line(0,1){58.6}}
        \put(17.575,17.575){\line(1,3){31.1}}
        \put(35.5,120){\line(3,-2){66.3}}
        \put(35.5,17.575){\line(4,1){66.5}}
        \put(120,60){\line(-5,1){120}}
        \put(26.3,44){\line(4,3){76.8}}
        \put(14.1,81){\line(1,-3){21.15}}
        \put(84.85,0){\line(-2,3){68}}
        \put(35.15,0){\circle{10}}
        \put(35.15,120){\circle{10}}
        \put(84.85,0){\circle{10}}
        \put(84.85,120){\circle{10}}
        \put(0,35.15){\circle{10}}
        \put(0,84.85){\circle{10}}
        \put(120,35.15){\circle{10}}
        \put(120,84.85){\circle{10}}
        \put(60,0){\circle{10}}
        \put(60,120){\circle{10}}
        \put(0,60){\circle{10}}
        \put(120,60){\circle{10}}
        \put(17.575,17.575){\circle{10}}
        \put(16.575,101.425){\circle{10}}
        \put(102.425,17.575){\circle{10}}
        \put(103.425,101.425){\circle{10}}
        \put(60,23.5){\circle{10}}
        \put(45.8,58.4){\circle{10}}
        \put(26.3,44){\circle{10}}
        \put(35.15,17.575){\circle{10}}
        \put(14.2,81.6){\circle{10}}
        \put(102,75.6){\circle{10}}
        \put(102,34){\circle{10}}
        \put(49,111){\circle{10}}

      \end{picture}
    } 

  \end{picture}
  \caption{\hbox to2mm{\hfill}(a) Greechie diagram for OML G7s1;
\hbox to2mm{\hfill} (b) Greechie diagram for OML G7s2.
\label{fig-g7s}}
\end{figure}

For any particular $n$ there may exist lattices smaller than G$n$ for
which this property holds.  These can be more efficient,
computationally, for proving that an equation derived in $n$GO is weaker
than $n$-Go or independent from $(n-1)$-Go.  Based on a computer
scan of all (legless) OMLs
with 3-atom blocks (see footnote at the end of Section
\ref{sec:oml-eqs}), up to and including a block count of 12 along with
selected lattices with block counts up to 17, we obtained the
following results.  Lattice G3, with 34 nodes, is the smallest that
violates 3-Go.  (In OMLs with 3-atom blocks, the number of nodes is
twice the number of atoms, plus 2.)  The Peterson OML, with 32 nodes
(vs.~44 nodes in G4), is the smallest that violates 4-Go but not 3-Go.
(Fig.~\ref{fig-oag45m}) Lattice G5s, with 42 nodes (vs.~54 nodes in
G5), is the smallest that violates 5-Go but not 4-Go. (Also
Fig.~\ref{fig-oag45m}) Lattices G6s1 and G6s2, each with 44 nodes
(vs.~64 nodes in G6), are two of three smallest that violate 6-Go but
not 5-Go. (Fig.~\ref{fig-oag6}) Of these three, G6s1 is one of two
with 14 blocks, whereas G6s2 has 15 blocks. Lattices G7s1 and G7s2
(Fig.~\ref{fig-g7s}) are two of several  smallest we obtained to
violate 7-Go but not 6-Go. They both have 50 nodes and 16 and 17
blocks, respectively (vs.~74 nodes and 21 blocks in G7).
We made use of a dynamic programming algorithm to obtain a
program for checking on $n$-Go which is so fast that no reasonable $n$
is a problem.  For example, to find G7s1 among 207767 Greechie diagrams
with 24 atoms and 16 blocks took an 800$\>$MHz PC less than two hours.

\begin{figure}[htbp]\centering
  \setlength{\unitlength}{1pt}
  \begin{picture}(240,120)(0,0)

    \put(5,13) { 
      \begin{picture}(124,110)(0,0) 
        \put(32.2,0){\line(1,0){55.6}}
        \put(32.2,100){\line(1,0){55.6}}
        \put(2.3,50){\line(3,5){29.9}}
        \put(117.7,50){\line(-3,5){29.9}}
        \put(2.3,50){\line(3,-5){29.9}}
        \put(117.7,50){\line(-3,-5){29.9}}
        \put(60,100){\line(0,-1){100}}
        \put(17.25,25){\line(5,3){84.8}}
        \put(102.75,25){\line(-5,3){84.8}}
        \put(34.71,65.6){\line(1,0){50.58}}

        \put(2.3,50){\circle{10}}
        \put(117.7,50){\circle{10}}
        \put(87.8,0){\circle{10}}
        \put(87.8,100){\circle{10}}
        \put(32.2,0){\circle{10}}
        \put(32.2,100){\circle{10}}
        \put(60,0){\circle{10}}
        \put(60,100){\circle{10}}
        \put(17.25,25){\circle{10}}
        \put(17.85,76){\circle{10}}
        \put(102.75,25){\circle{10}}
        \put(102.15,76){\circle{10}}
        \put(34.71,65.6){\circle{10}}
        \put(85.29,65.6){\circle{10}}
        \put(60,65.6){\circle{10}}
      \end{picture}
    } 

    \put(170,0) { 
      \begin{picture}(124,110)(0,0) 
        \put(35.15,0){\line(1,0){49.7}}
        \put(35.15,120){\line(1,0){49.7}}
        \put(0,35.15){\line(0,1){49.7}}
        \put(120,35.15){\line(0,1){49.7}}
        \put(0,35.15){\line(1,-1){35.15}}
        \put(0,84.85){\line(1,1){35.15}}
        \put(120,35.15){\line(-1,-1){35.15}}
        \put(120,84.85){\line(-1,1){35.15}}
        \put(60,120){\line(0,-1){71}}
        \put(0,60){\line(5,2){103.5}}
        \put(120,60){\line(-5,2){103.5}}
        \put(17.575,17.575){\line(4,3){72.5}}
        \put(102.425,17.575){\line(-4,3){72.5}}

        \put(35.15,0){\circle{10}}
        \put(35.15,120){\circle{10}}
        \put(84.85,0){\circle{10}}
        \put(84.85,120){\circle{10}}
        \put(0,35.15){\circle{10}}
        \put(0,84.85){\circle{10}}
        \put(120,35.15){\circle{10}}
        \put(120,84.85){\circle{10}}
        \put(60,0){\circle{10}}
        \put(60,120){\circle{10}}
        \put(0,60){\circle{10}}
        \put(120,60){\circle{10}}
        \put(17.575,17.575){\circle{10}}
        \put(16.575,101.425){\circle{10}}
        \put(102.425,17.575){\circle{10}}
        \put(103.425,101.425){\circle{10}}
        \put(60,101.425){\circle{10}}
        \put(60,49){\circle{10}}
        \put(30.3,72){\circle{10}}
        \put(89.7,72){\circle{10}}

      \end{picture}
    } 

  \end{picture}
  \caption{\hbox to3mm{\hfill}(a)
    Peterson OML;
  \hbox to2mm{\hfill} (b) Greechie diagram for OML G5s.
\label{fig-oag45m}}
\end{figure}
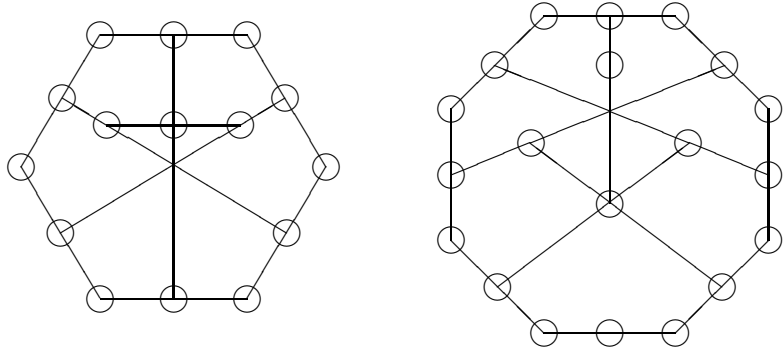

\begin{figure}[htbp]\centering
  \setlength{\unitlength}{1pt}
  \begin{picture}(240,120)(0,0)

    \put(-40,0) { 
      \begin{picture}(124,110)(0,0) 
        \put(35.15,0){\line(1,0){49.7}}
        \put(35.15,120){\line(1,0){49.7}}
        \put(0,35.15){\line(0,1){49.7}}
        \put(120,35.15){\line(0,1){49.7}}
        \put(0,35.15){\line(1,-1){35.15}}
        \put(0,84.85){\line(1,1){35.15}}
        \put(120,35.15){\line(-1,-1){35.15}}
        \put(120,84.85){\line(-1,1){35.15}}
        \put(60,120){\line(0,-1){120}}
        \put(0,60){\line(5,2){103.5}}
        \put(120,60){\line(-5,2){103.5}}
        \put(17,18.2){\line(2,3){67.8}}
        \put(103,18.2){\line(-2,3){67.8}}
        \put(60,0){\line(-1,4){23.3}}

        \put(35.15,0){\circle{10}}
        \put(35.15,120){\circle{10}}
        \put(84.85,0){\circle{10}}
        \put(84.85,120){\circle{10}}
        \put(0,35.15){\circle{10}}
        \put(0,84.85){\circle{10}}
        \put(120,35.15){\circle{10}}
        \put(120,84.85){\circle{10}}
        \put(60,0){\circle{10}}
        \put(60,120){\circle{10}}
        \put(0,60){\circle{10}}
        \put(120,60){\circle{10}}
        \put(17,18.2){\circle{10}}
        \put(16.575,101.425){\circle{10}}
        \put(103,18.2){\circle{10}}
        \put(103.425,101.425){\circle{10}}
        \put(60,38){\circle{10}}
        \put(40.8,76.2){\circle{10}}
        \put(36.7,93.2){\circle{10}}
        \put(30,38){\circle{10}}
        \put(90,38){\circle{10}}

      \end{picture}
    } 

    \put(170,0) { 
      \begin{picture}(124,110)(0,0) 
        \put(39.48,3.62){\line(1,0){41.04}}
        \put(39.48,3.62){\line(-6,5){31.44}}
        \put(80.52,3.62){\line(6,5){31.44}}
        \put(8.04,30){\line(-1,6){6.75}}
        \put(111.96,30){\line(1,6){6.75}}
        \put(0.91,70.42){\line(3,5){21.2}}
        \put(119.09,70.42){\line(-3,5){21.2}}
        \put(22.13,105.96){\line(3,1){38}}
        \put(97.87,105.96){\line(-3,1){38}}
        \put(60,3.62){\line(0,1){114.98}}
        \put(40.715,112.3){\line(3,-5){56.7}}
        \put(79.285,112.3){\line(-3,-5){56.7}}
        \put(8.04,30){\line(5,3){99.5}}
        \put(111.96,30){\line(-5,3){99.5}}
        \put(4.475,50.21){\line(1,0){111.05}}

        \put(80.52,3.62){\circle{10}}
        \put(39.48,3.62){\circle{10}}
        \put(111.96,30){\circle{10}}
        \put(8.04,30){\circle{10}}
        \put(119.09,70.42){\circle{10}}
        \put(0.91,70.42){\circle{10}}
        \put(97.2,105.8){\circle{10}}
        \put(21.8,105.8){\circle{10}}
        \put(60,118.6){\circle{10}}
        \put(60,3.62){\circle{10}}
        \put(97.24,17.81){\circle{10}}
        \put(22.76,17.81){\circle{10}}
        \put(115.525,50.21){\circle{10}}
        \put(4.475,50.21){\circle{10}}
        \put(107.83,89.19){\circle{10}}
        \put(12.17,89.19){\circle{10}}
        \put(79.285,112.5){\circle{10}}
        \put(40.715,112.5){\circle{10}}
        \put(60,50.21){\circle{10}}
        \put(51.3,66.6){\circle{10}}
        \put(68.7,66.6){\circle{10}}

      \end{picture}
    } 

  \end{picture}
  \caption{\hbox to3mm{\hfill}(a)
    Greechie diagram for OML G6s1;
  \hbox to2mm{\hfill} (b) Greechie diagram for OML G6s2.
\label{fig-oag6}}
\end{figure}

The next lemma provides some technical results for subsequent use.
Note that $a\perp b\perp c$ means $a\perp b$ and $b\perp c$ (but not
necessarily $a\perp c$).
\begin{lemma}In any {\em OML} we have:
\begin{eqnarray}
&a\perp b\perp c\qquad\Rightarrow\qquad
     (a\cup b)\cap(a\to_2 c)\le b\cup c\label{eq:govar}\\
&a\perp b\perp c\qquad\Rightarrow\qquad a\cup b\le c\to_2 a\label{eq:govar2}\\
&a\perp b\perp c\qquad\&\qquad (c\to_2 a)\cap d\le a\to_2 c
     \qquad\Rightarrow\qquad
     (a\cup b)\cap d\le b\cup c\label{eq:gon2n}
\end{eqnarray}
\end{lemma}
\begin{proof}
For (\ref{eq:govar}):  $(a\cup
b)\cap(a\to_2 c) = (a\cup b)\cap(c\cup(a'\cap c'))$.  {}From hypotheses,
$b$ commutes with $a$ and $c\cup(a'\cap c')$.  Using F-H twice,
$(a\cup b)\cap(c\cup(a'\cap c'))=
(b\cap(c\cup(a'\cap c')))\cup(a\cap c)\cup(a\cap a'\cap c')\le b\cup c$.
For (\ref{eq:govar2}):  {}From hypotheses, $a\cup b\le a\cup(c'\cap
a')=c\to_2 a$.  For (\ref{eq:gon2n}):  {}From (\ref{eq:govar2}), $(a\cup
b)\cap d\le (c\to_2 a)\cap d\le$[from hypothesis] $a\to_2 c$.  Thus
$(a\cup b)\cap d\le (a\cup b)\cap(a\to_2 c)$, which by
(\ref{eq:govar}) is $\le b\cup c$.
\end{proof}

The $n$-Go equations can be equivalently expressed as inferences
involving $2n$ variables, as the following theorem shows.  In this form
they can be useful for certain kinds of proofs, as we
illustrate in Theorem~\ref{th:mayet}.
\begin{theorem}\label{th:go2n} Any {\em OML} in which
\begin{eqnarray}
\lefteqn{a_1\perp b_1\perp a_2\perp b_2\perp\ldots a_n\perp b_n\perp a_1
    \qquad\Rightarrow} & & \nonumber \\
& & (a_1\cup b_1)\cap (a_2\cup b_2)\cap\cdots(a_n\cup b_n)\le b_1\cup a_2
\label{eq:go2n}
\end{eqnarray}
holds is an {\em $n$GO} and vice versa.
\end{theorem}
\begin{proof}
Substituting $c_1$ for $a_1$,\ldots,
$c_n$ for $a_n$, $c_1'\cap c_2'$ for $b_1$,
\ldots, $c_{n-1}'\cap c_n'$ for $b_{n-1}$, and $c_n'\cap c_1'$ for $b_n$, we
satisfy the hypotheses of (\ref{eq:go2n}) and
obtain (\ref{eq:godow2d}).

Conversely, suppose the hypotheses of (\ref{eq:go2n}) hold.
{}From the hypotheses and (\ref{eq:govar2}), we obtain
$(a_2\cup b_2)\cdots\cap
(a_{n-1}\cup b_{n-1})\cap(a_n\cup b_n)\le(a_3\to_2 a_2)\cdots\cap
(a_n\to_2 a_{n-1})\cap(a_1\to_2 a_n)$.
Thus
$(a_2\to_2 a_1)\cap[(a_2\cup b_2)\cdots\cap
(a_{n-1}\cup b_{n-1})\cap(a_n\cup b_n)]\le
(a_2\to_2 a_1)\cap[(a_3\to_2 a_2)\cdots\cap
(a_n\to_2 a_{n-1})\cap(a_1\to_2 a_n)]=
(a_2\to_2 a_1)\cap(a_1\to_2 a_n)\cap(a_n\to_2 a_{n-1})\cdots\cap
(a_3\to_2 a_2)$.
Applying (\ref{eq:godow2d}) to the right-hand side, we obtain
$(a_2\to_2 a_1)\cap[(a_2\cup b_2)\cdots\cap
(a_{n-1}\cup b_{n-1})\cap(a_n\cup b_n)]\le
a_1\to_2 a_2$.
Then (\ref{eq:gon2n}) gives us (\ref{eq:go2n}).
\end{proof}

Mayet \cite{mayet86} presents a method for obtaining equations that hold
in all lattices with a strong or full set of states.  However, it
turns out that the examples of those equations he shows are implied by
the $n$-Go equations and thus do not provide us with additional
information about lattices with strong states or ${\cal C}({\cal H})$ in
particular.  To the authors' knowledge, there is no known example of
such an equation that cannot be derived from the $n$-Go equations.
It apparently remains an open problem whether Mayet's method gives
equations  that hold in all OMLs with a strong set of states but that
cannot be derived from equations $n$-Go.
\begin{theorem}\label{th:mayet}
The following equations (derived as Examples~2, 3, and
4 in \cite{mayet86})
hold in {\em 3GO}, {\em 6GO}, and {\em 4GO} respectively.
\begin{eqnarray}
\lefteqn{\qquad\qquad\qquad\qquad(a\to_1 b)\cap(b\to_1 c)
\cap(c\to_1 a)\le b\to_1 a} & &
\label{eq:mayet2} \\
\lefteqn{a\perp b\perp c\perp d\perp e\perp f\perp a
\qquad \Rightarrow}& & \nonumber \\
& & (a\cup b)\cap(d\cup e)'\cap((((a\cup b)\to_1(d\cup e)'
)\to_1((e\cup f)\to_1(b\cup c)')')'\to_1(c\cup d)) \nonumber \\
\lefteqn{\qquad \le b\cup c\cup(e\cup f)'} & &
\label{eq:mayet3} \\
\lefteqn{a\perp b\perp c\perp d\perp e\perp f\perp g\perp h\perp a
\qquad \Rightarrow}& & \nonumber \\
\lefteqn{\qquad (a\cup b)\cap(c\cup d)\cap(e\cup f)\cap(g\cup h)
\cap((a\cup h)\to_1(d\cup e)') = 0} & &
\label{eq:mayet4}
\end{eqnarray}
\end{theorem}
\begin{proof}
For (\ref{eq:mayet2}): This is the same as (\ref{eq:godow1d}) for
$n=3$.

For (\ref{eq:mayet3}):  Using (\ref{eq:go2n}) we
express the 6GO law as
\begin{eqnarray}
\lefteqn{a_1\perp b_1\perp a_2\perp b_2\perp a_3\perp b_3\perp a_4\perp
    b_4\perp a_5\perp b_5\perp a_6\perp b_6\perp a_1
    \qquad\Rightarrow} & & \nonumber \\
& & (a_1\cup b_1)\cap(a_2\cup b_2)\cap(a_3\cup b_3)\cap(a_4\cup b_4)
        \cap(a_5\cup b_5)\cap(a_6\cup b_6)\le b_1\cup a_2
\,.
\label{eq:mayet3proof1}
\end{eqnarray}
We define $p=((a\cup b)\to_1(d\cup e)')'$, $q=((e\cup f)\to_1(b\cup
c)')'$, and $r=(p'\to_1 q)'\cap(c\cup d)$.  In (\ref{eq:mayet3proof1})
we substitute $a$ for $a_1$, $b$ for $b_1$, $c$ for $a_2$, $(c\cup d)'$
for $b_2$, $r$ for $a_3$, $p'\to_1 q$ for $b_3$, $(p'\to_1 q)'$ for
$a_4$, $p'\cap q$ for $b_4$, $q'$ for $a_5$, $q$ for $b_5$, $(e\cup f)'$
for $a_6$, and $f$ for $b_6$.  With this substitution, all hypotheses of
(\ref{eq:mayet3proof1}) are satisfied by the hypotheses of
(\ref{eq:mayet3}).  The conclusion becomes
\begin{eqnarray}
\lefteqn{(a\cup b)\cap(c\cup(c\cup d)')\cap(r\cup(p'\to_1 q))}
\nonumber \\
& & \cap((p'\to_1 q)'\cup
(p'\cap q))\cap(q'\cup q)\cap((e\cup f)'\cup f) \le b\cup c
\,.
\label{eq:mayet3proof2}
\end{eqnarray}
We simplify (\ref{eq:mayet3proof2}) using $c\cup(c\cup d)'=$[since $c$
and $d$ commute by hypothesis] $(c\cup c')\cap(c\cup d')=1\cap(c\cup
d')=$[since $c\le d'$] $d'$; $(e\cup f)'\cup f=e'$ similarly; $(p'\to_1
q)'\cup(p'\cap q)=p'$; and $q'\cup q=1$.  This gives us
\begin{eqnarray}
(a\cup b)\cap d'\cap(r\cup(p'\to_1 q))\cap p'\cap e' & \le & b\cup c
\,.
\label{eq:mayet3proof3}
\end{eqnarray}
Now, in any OML we have $p'=(a\cup b)\to_1(d\cup e)'=(a\cup b)'\cup
((a\cup b)\cap(d\cup e)')\ge (a\cup b)\cap(d\cup e)'
\ge (a\cup b)\cap(d\cup e)'\cap((p'\to_1 q)\cup r)$.  Thus the left-hand
side of (\ref{eq:mayet3proof3}) absorbs $p'$, so
\begin{eqnarray}
(a\cup b)\cap d'\cap(r\cup(p'\to_1 q))\cap e' & \le & b\cup c
  \label{eq:fornew3go} \\
     & \le & b\cup c
\cup(e\cup f)'   \nonumber
\end{eqnarray}
which after rearranging is exactly (\ref{eq:mayet3}).

For (\ref{eq:mayet4}): Using (\ref{eq:go2n}) we obtain
from the 4GO law
\begin{eqnarray}
\lefteqn{a\perp b\perp c\perp d\perp e\perp f\perp g\perp
    h\perp a
    \qquad\Rightarrow} & & \nonumber \\
& & (a\cup b)\cap(c\cup d)\cap(e\cup f)\cap(g\cup h)
\le(a\cup h)\cap(d\cup e)
\,. \nonumber
\end{eqnarray}
Therefore
\begin{eqnarray}
\lefteqn{a\perp b\perp c\perp d\perp e\perp f\perp g\perp
    h\perp a
    \qquad\Rightarrow} & & \nonumber \\
& & (a\cup b)\cap(c\cup d)\cap(e\cup f)\cap(g\cup h)
\cap((a\to h)\to_1(d\cup e)') \nonumber \\
& & \le(a\cup h)\cap(d\cup e)\cap((a\cup h)\to_1(d\cup e)')
\,. \nonumber
\end{eqnarray}
In any OML we have $x\cap y\cap(x\to_1 y')=0$; applying this
to the right-hand side we obtain (\ref{eq:mayet4}).
\end{proof}

To the authors' knowledge all 3-variable equations published so far that
hold in all OMLs with a strong set of states are derivable in 3GO.
Below we show an equation with 3 variables that is derivable in 6GO but
is independent from the 3GO law.  It shows that it is possible to
express with only 3 variables a property that holds only in $n$GOs
smaller than 3GO.
\medskip\noindent
\begin{theorem}\label{th:new3go}
The 3-variable equation
\begin{eqnarray}
\lefteqn{((a \to_2 b) \cap (a \to_2 c)') \cap
 (( ((a \to_2 b) \to_1 (a \to_2 c)')
\to_1 ((b \to_2 c) \to_1 (b \to_2 a)')' )'}\qquad\qquad\qquad
\qquad\qquad\qquad\qquad\qquad\qquad\qquad \nonumber\\
& &  \to_1 (c \to_2 a)) \le b \to_2 a
\label{eq:new3go}
\end{eqnarray}
holds in a {\rm 6GO} but cannot be derived (in an {\em OML}) from the
{\em 3GO} law nor vice versa.
\end{theorem}
\begin{proof}
To show this equation holds in 6GO, we start with (\ref{eq:fornew3go})
that occurs in the proof of Mayet's Example 3, rewriting
it as:
\begin{eqnarray}
\lefteqn{d\perp e\perp f\perp g\perp h\perp j\perp d
\qquad \Rightarrow}& & \nonumber \\
& & (d\cup e)\cap(g\cup h)'\cap((((d\cup e)\to_1(g\cup h)'
)\to_1((h\cup j)\to_1(e\cup f)')')'\to_1(f\cup g)) \nonumber \\
\lefteqn{\qquad \le e\cup f} & &
\label{eq:mayet3truncated}
\end{eqnarray}
We substitute $b$ for $d$, $a'\cap b'$ for $e$, $a$ for $f$, $a'\cap c'$
for $g$, $c$ for $h$, and $c'\cap b'$ for $j$.  With these substitutions,
the hypotheses of (\ref{eq:mayet3truncated}) are satisfied.  This results
in (\ref{eq:new3go}), showing that (\ref{eq:new3go}) holds in 6GO.

We show independence as follows.  On the one hand, (\ref{eq:new3go})
fails in the Peterson OML (Fig.~\ref{fig-oag45m}a) but holds in
OML G3 (Fig.~\ref{fig-oag34}a).  On the other hand, the 3GO law
(\ref{eq:godow3o}) holds in the Peterson OML but fails in
G3.
\end{proof}
It is not known whether (\ref{eq:new3go}) holds in 4GO or 5GO.

Using our results so far we can show that
$a_i{\buildrel\gamma\over\equiv}a_j=1$ is similar to a relation of
equivalence (although strictly speaking it is not one, since
$a_i{\buildrel\gamma\over\equiv}a_j$ involves not 2 but
$|j-i|+1$ variables).
Reflexivity $a_i{\buildrel\gamma\over\equiv}a_i=1$ follows by
definition, symmetry $a_i{\buildrel\gamma\over\equiv}a_j=1 \Rightarrow
a_j{\buildrel\gamma\over\equiv}a_i=1$ from the Godowski equations, and
transitivity $a_i{\buildrel\gamma\over\equiv}a_j=1
\ \&\ a_j{\buildrel\gamma\over\equiv}a_k=1 \Rightarrow
a_i{\buildrel\gamma\over\equiv}a_k=1$ from the
following theorem. Analogous results can be stated for
${\buildrel\delta\over\equiv}$. An open problem is whether there
exists an equation corresponding to
$a_i{\buildrel\gamma\over\equiv}a_j=1$ and
$a_i{\buildrel\delta\over\equiv}a_j=1$ as in Equations
(\ref{eq:qm-as-id}), (\ref{eq:id/c1}), (\ref{eq:id/c}), and
(\ref{eq:id/cd}).

\begin{theorem}\label{th:god-trans}
The following equation holds in {\em $n$GO}, where $i,j\ge 1$ and
$n=\max(i,j,3)$.
\begin{eqnarray}
(a_1{\buildrel\gamma\over\equiv}a_i)
\cap(a_i{\buildrel\gamma\over\equiv}a_j)
&\le&a_1{\buildrel\gamma\over\equiv}a_j
\end{eqnarray}
\end{theorem}
\begin{proof}
If ${\buildrel\gamma\over\equiv}$ has 3 or more variables, we replace it
with a chained identity per (\ref{eq:godow1c}), otherwise we replace it
with the extended definition we mention after Definition
\ref{def:god-equiv}.  The proof is then obvious.  (In many cases the
equation may also hold for smaller $n$ or even in OML or OL, e.g.\ when
$j=1$.)
\end{proof}

The next lemma shows an interesting ``variable-swapping'' property of the
Godowski identity that we shall use in a later proof
[of Theorem \ref{th:godistr}].

\begin{lemma}\label{lem:goswap}
In any {\em OML} we have
\begin{eqnarray}
(a_1{\buildrel\gamma\over\equiv}a_n)\cap a_i' =
 a_1' \cap a_2' \cdots\cap a_n'\,,\qquad i=1,\ldots,n\,.
\label{eq:goswap}
\end{eqnarray}
In particular,
\begin{eqnarray}
(a_1{\buildrel\gamma\over\equiv}a_n)\cap a_i'=
(a_1{\buildrel\gamma\over\equiv}a_n)\cap a_j'\,,\qquad
i=1,\ldots,n\,,\quad j=1,\ldots,n\,.
\label{eq:goswap2}
\end{eqnarray}
\end{lemma}
\begin{proof}
We illustrate the case $i=1$. In any OML we have
$a'\cap(b\to_1 a)=a'\cap b'$.  Thus
$(a_1\to_1a_2)\cdots\cap(a_{n-2}\to_1a_{n-1})\cap
  (a_{n-1}\to_1a_n)\cap(a_n\to_1a_1)\cap a_1'=
(a_1\to_1a_2)\cdots\cap(a_{n-2}\to_1a_{n-1})\cap
  (a_{n-1}\to_1a_n)\cap a_n'\cap a_1'=\cdots=
a_1'\cap a_2'\cdots\cap a_{n-1}'\cap a_n'\cap a_1'$.
\end{proof}

\section{\large Orthoarguesian Equations}
\label{sec:oa}

In this section we show that all orthoarguesian-based equalities (which
must hold in any Hilbert lattice) that have appeared in the literature as
equations with 4 and 6 variables can be reduced to just two
equations with 3 and 4 variables. The latter two equations we call
the 3OA and 4OA laws, respectively, and introduce them by Definition
\ref{def:3OA&4OA}. Their equivalence to the afore mentioned 4- and
6-variables equations is shown in Theorems \ref{th:go-gr3oa}
and \ref{th:4oa-go-gr} and in Theorem \ref{th:6oa}, respectively.
A new 3-variable consequence of the 4OA law which is not equivalent
to the 3OA
law is given by Theorem \ref{th:new3oa}. Possibly equivalent
inference forms of the 3OA law and the 4OA law are given by Theorems
\ref{th:oa-eq-c-3}, \ref{th:oa-eq-c-4}, and \ref{th:oa-equiv}.

\begin{definition}\label{def:3-4-oa}\hfil
\begin{eqnarray}
a{\buildrel c\over\equiv_i}b\
&{\buildrel\rm def\over =}&\ ((a\to_i c)
\cap(b\to_i c))
\cup((a'\to_i c)\cap(b'\to_i c)),
\qquad i=1,3,\\
a{\buildrel c\over\equiv_i}b\
&{\buildrel\rm def\over =}&\ ((c\to_i a)
\cap(c\to_i b))
\cup((c\to_i a')\cap(c\to_i b')),\qquad i=2,4,\\
a{\buildrel c,d\over\equiv_i}b\
&{\buildrel\rm def\over =}&\ (a{\buildrel d\over\equiv_i}b)\cup
((a{\buildrel d\over\equiv_i}c)\cap
(b{\buildrel d\over\equiv_i}c)),\hspace{1.7in} i=1,\dots,4.
\end{eqnarray}
We call $a{\buildrel c\over\equiv_i}b$ a 3-variable orthoarguesian
identity and $a{\buildrel c,d\over\equiv_i}b$ a 4-variable
orthoarguesian identity and denote them as {\em 3-oa} and {\em 4-oa}
respectively.
\end{definition}

\begin{theorem}\label{th:oa-eq-c-3} An ortholattice to which
any of
\begin{eqnarray}
a{\buildrel c\over\equiv_i}b=1\qquad &\Leftrightarrow &\qquad
a\to_ic=b\to_ic,\qquad i=1,3\label{eq:id/c1}\\
a{\buildrel c\over\equiv_i}b=1\qquad &\Leftrightarrow &\qquad
c\to_ia=c\to_ib,\qquad i=2,4\,
\label{eq:id/c}
\end{eqnarray}
are added is a variety smaller than {\em OML} that fails in
lattice {\em L28} (Fig.~\ref{fig-oa}a).
\end{theorem}

The corresponding expressions for $i=5$ do not hold in a Hilbert
lattice (right to left implications fail in MO2).

\begin{figure}[htbp]\centering
  \setlength{\unitlength}{1pt}
  \begin{picture}(360,150)(30,-10)

    \put(80,13) { 
      \begin{picture}(124,120)(0,0) 
        \put(32.2,0){\line(1,0){55.6}}
        \put(32.2,100){\line(1,0){55.6}}
        \put(2.3,50){\line(3,5){29.9}}
        \put(117.7,50){\line(-3,5){29.9}}
        \put(2.3,50){\line(3,-5){29.9}}
        \put(117.7,50){\line(-3,-5){29.9}}
        \put(60,100){\line(0,-1){100}}
        \put(2.3,50){\circle{10}}
        \put(117.7,50){\circle{10}}
        \put(87.8,0){\circle{10}}
        \put(87.8,100){\circle{10}}
        \put(32.2,0){\circle{10}}
        \put(32.2,100){\circle{10}}
        \put(60,0){\circle{10}}
        \put(60,100){\circle{10}}
        \put(17.25,25){\circle{10}}
        \put(17.25,75){\circle{10}}
        \put(102.75,25){\circle{10}}
        \put(102.75,75){\circle{10}}
        \put(60,50){\circle{10}}
      \end{picture}
    } 

    \put(290,0) { 

      \begin{picture}(124,120)(0,0) 
        \put(110.9,96.2){\line(-2,1){49.3}}
        \put(110.9,96.2){\line(1,-4){13}}
        \put(89,0){\line(4,5){35}}
        \put(89,0){\line(-1,0){54.8}}
        \put(34.2,0){\line(-4,5){35}}
        \put(12.2,96.2){\line(-1,-4){13}}
        \put(12.2,96.2){\line(2,1){49.3}}
        \put(61.5,0){\line(0,1){120.5}}
        \put(12.5,96.2){\line(5,-4){93.5}}
        \put(110.5,96.2){\line(-5,-4){93.5}}

        \put(61.5,120.5){\circle{10}}
        \put(110.5,96.2){\circle{10}}
        \put(123.5,44){\circle{10}}
        \put(89,0){\circle{10}}
        \put(34,0){\circle{10}}
        \put(-0.5,44){\circle{10}}
        \put(12.5,96.2){\circle{10}}
        \put(86.2,108.4){\circle{10}}
        \put(117.5,70){\circle{10}}
        \put(106.0,21.4){\circle{10}}
        \put(61.5,0){\circle{10}}
        \put(17.1,21.4){\circle{10}}
        \put(5.5,70){\circle{10}}
        \put(36.9,108.4){\circle{10}}
        \put(61.5,90.4){\circle{10}}
        \put(42,41){\circle{10}}
        \put(81,41){\circle{10}}
      \end{picture}
    } 

  \end{picture}
  \caption{\hbox to3mm{\hfill}(a)
   Greechie diagram for OML L28;
  \hbox to1cm{\hfill} (b) Greechie diagram for OML L36.
\label{fig-oa}}
\end{figure}

\begin{theorem}\label{th:oa-eq-c-4} An ortholattice to which
any of
\begin{eqnarray}
a{\buildrel c,d\over\equiv_i}b=1\qquad \Leftrightarrow\qquad
a\to_id=b\to_id,\qquad i=1,3
\label{eq:id/cd}
\end{eqnarray}
is added is a variety smaller than {\em OML} that fails in lattice
{\em L36} (Fig.~\ref{fig-oa}b) for $i=1,3$.
\end{theorem}

The new identities ${\buildrel c\over\equiv_1}$ and
${\buildrel c,d\over\equiv_1}$ being equal to one are relations of
equivalence. It is obvious that they are reflexive ($a{\buildrel
c\over\equiv_1}a=1$, $a{\buildrel c,d\over\equiv_1}a=1$), and
symmetric ($a{\buildrel c\over\equiv_1}b=1\Rightarrow b{\buildrel
c\over\equiv_1}a=1$, \ $a{\buildrel c,d\over\equiv_1}b=1\Rightarrow
b{\buildrel c,d\over\equiv_1}a=1$), and the
transitivity follows from Theorem \ref{th:oa-equiv} below.
They are, however, not relations of congruence because
$a\>{\buildrel c\over\equiv_1}\>b=1\>\Rightarrow (a\cup d)
\>{\buildrel c\over\equiv_1}\>(b\cup d)=1$ does not hold: it fails
in the {\it Chinese lantern} MO2 (Fig.~\ref{fig:o6mo2}b). Conditions
(\ref{eq:id/c1}), (\ref{eq:id/c}) and (\ref{eq:id/cd}) must hold in
any Hilbert space (and therefore by any quantum
simulator) for $i=1$ as we show below. Expressions corresponding to
Eq.~(\ref{eq:id/cd}) for $i=2,4,5$
do not hold in a Hilbert lattice and it is an open problem whether
there exist equivalent relations of equivalence for $i=2,4,5$.
In what follows we keep to $i=1$ (and not $i=3$) because $i=1$ enables
us to switch to the Sasaki projection $\varphi_ab=(a\to_1b')'$
of $b$ on $a$ later on. The Sasaki projection plays an important role in
the definition of the covering property which is a consequence of the
superposition principle. \cite{beltr-cass-book}

\begin{definition}\label{def:3OA&4OA} Let $a{\buildrel c\over\equiv}b\
{\buildrel\rm def\over =}\ a{\buildrel c\over\equiv_1}b$ \ and
\ $a{\buildrel c,d\over\equiv}b\ {\buildrel\rm def\over =}\
a{\buildrel c,d\over\equiv_1}b$.

A {\em 3OA} is an {\em OL} in which the
following additional condition is satisfied:
\begin{eqnarray}
(a\to_1 c) \cap (a{\buildrel c\over\equiv}b)
\le b\to_1 c\,.\label{eq:3oa}
\end{eqnarray}
A {\em 4OA} is an {\em OL} in which the
following additional condition is satisfied:
\begin{eqnarray}
(a\to_1 d) \cap (a{\buildrel c,d\over\equiv}b)
\le b\to_1 d\,.\label{eq:4oa}
\end{eqnarray}
\end{definition}

Note that the 3OA and 4OA laws (\ref{eq:3oa}) and (\ref{eq:4oa})
have 3 and 4 variables respectively. Both 3OA and 4OA laws fail in
O6, so, they are OMLs, but there exist OMLs that are
neither 3OAs nor 4OAs: equations (\ref{eq:3oa}) and (\ref{eq:4oa})
both fail in the orthomodular lattice
L28 (Fig.~\ref{fig-oa}a).

\medskip\noindent
\begin{theorem}
Every {\em 4OA} is a {\em 3OA}, but there exist {\em 3OA}s that are not
{\em 4OA}s.
\end{theorem}
\begin{proof}
In $(a\to_1 d) \cap (a{\buildrel c,d\over\equiv}b) \le (b\to_1 d)$,
set $c=b$.  On the other
hand, lattice L36 (Fig.~\ref{fig-oa}b) is a 3OA because it is an OML in
which (\ref{eq:3oa}) holds, but it is not a 4OA because it
violates (\ref{eq:4oa}).
\end{proof}

The next lemma provides some technical results for use in subsequent
proofs.

\begin{lemma}In any {\em OML} we have:
\begin{eqnarray}
&(a\to_1 b)\cap a=a\cap b\label{eq:oalem1}\\
&(a\to_1 b)\cap(a'\to_1 b)=(a\to_1 b)\cap b=(a\cap b)\cup (a'\cap b)
 \label{eq:oalem2}\\
&(a'\to_1 b)'\le a'\le a\to_1 b \label{eq:oalem3}\\
&(a\to_1 b)\to_1 b=a'\to_1 b \label{eq:oalem4}\\
&(a\to_1 b)'\to_1 b=a\to_1 b \label{eq:oalem5}\\
&(a\to_i b)\cup (a\to_j b)=a\to_0 b,\ i,j=0,\ldots ,4,
\ i\ne j\label{eq:oalem6}\\
&a'\le b \qquad\Rightarrow\qquad b\le a\to_1 b \label{eq:oalem8}\\
&a\cap((a\to_1 c)\cup b)\le c \qquad\Leftrightarrow\qquad
   b\le a\to_1 c \label{eq:oalem9}
\end{eqnarray}
\end{lemma}
\begin{proof}
For (\ref{eq:oalem1})--(\ref{eq:oalem8}): We omit the easy proofs.

For (\ref{eq:oalem9}):
If $a\cap((a\to_1 c)\cup b)\le c$ then
$a\cap((a\to_1 c)\cup b)\le a\cap c = (a\to_1 c)\cap a$ using
(\ref{eq:oalem1}), so $b\le 1\cap((a\to_1 c)\cup b)=
((a\to_1 c)\cup a)\cap((a\to_1 c)\cup((a\to_1 c)\cup b))=$ (via
F-H) $(a
\to_1 c)\cup(a\cap((a\to_1 c)\cup b))\le (a\to_1 c)\cup((a\to_1 c)\cap a)=
a\to_1 c$.  Conversely, if $b\le a\to_1 c$, then
using (\ref{eq:oalem1}),
$a\cap((a\to_1 c)\cup b)= a\cap(a\to_1 c)=a\cap c\le c$.
%
%
\end{proof}

In the next theorem we show that the 4OA law (\ref{eq:4oa}) is
equivalent to the orthoarguesian law (\ref{eq:6oa}) discovered by A.~Day
(cf.~\cite{go-gr,gr-non-s}), which holds in ${\cal C}({\cal H})$.  Thus
the 4OA law also holds in ${\cal C}({\cal H})$.

\medskip\noindent
\begin{theorem}\label{th:6oa} An {\em OML} in which
\begin{eqnarray}
\lefteqn{a \perp b \qquad \&\qquad c \perp d \qquad \&\qquad e
  \perp f \qquad \Rightarrow} & & \nonumber \\
& & ( a \cup b ) \cap ( c \cup d ) \cap ( e \cup f ) \le \nonumber \\
& &b \cup ( a \cap ( c \cup ( ( ( a \cup c )
\cap ( b \cup d ) ) \cap
( ( ( a \cup e ) \cap ( b \cup f ) ) \cup
( ( c \cup e ) \cap ( d \cup f ) ) ) ) ) )\quad \label{eq:6oa}
\end{eqnarray}
(where $a\perp b\ {\buildrel\rm def\over =}\ a\le b'$)
holds is a {\em 4OA} and vice versa.
\end{theorem}
\begin{proof}
We will work with the dual of (\ref{eq:6oa}),
\begin{eqnarray}
\lefteqn{a' \le b \qquad \&\qquad c' \le d \qquad \&\qquad e'
  \le f \qquad \Rightarrow} & & \nonumber \\
& &b \cap ( a \cup ( c \cap ( ( ( a \cap c )
\cup ( b \cap d ) ) \cup
( ( ( a \cap e ) \cup ( b \cap f ) ) \cap
( ( c \cap e ) \cup ( d \cap f ) ) ) ) ) ) \le \nonumber \\
& & ( a \cap b ) \cup ( c \cap d ) \cup ( e \cap f )\,.  \label{eq:6oadual}
\end{eqnarray}
First we show that the 4OA law implies (\ref{eq:6oadual}).
In any OL we have
\begin{eqnarray}
\lefteqn{ b \le g \to_1 k \qquad \&\qquad d \le h \to_1 k \qquad \&\qquad
 f \le j \to_1 k \qquad \Rightarrow} & & \nonumber \\
& & b \cap ( g \cup ( h \cap ( ( ( g \cap h )
\cup ( b \cap d ) ) \cup
( ( ( g \cap j ) \cup ( b \cap f ) ) \cap
( ( h \cap j ) \cup ( d \cap f ) ) ) ) ) ) \le \nonumber \\
& &( g \to_1 k )\cap( g \cup( h \cap(
        (( g \cap h )\cup(( g \to_1 k )\cap( h \to_1 k )))\cup \nonumber \\
& &   ((( g \cap j )\cup(( g \to_1 k )\cap( j \to_1 k )))\cap
   (( h \cap j )\cup(( h \to_1 k )\cap( j \to_1 k )))))))\,.\label{eq:6oaproof4}
\end{eqnarray}
Substituting $ a '\to_1 k $ for $ g $, $ c'\to_1 k $ for $h$, and $e'\to_1 k $
for $j$; simplifying with (\ref{eq:oalem4}); and applying
(\ref{eq:oalem3}) to the left-hand side of the conclusion
we obtain
\begin{eqnarray}
\lefteqn{ b \le a \to_1 k \qquad \&\qquad d \le c \to_1 k \qquad \&\qquad
 f \le e \to_1 k \qquad \Rightarrow} & & \nonumber \\
& &b \cap ( a \cup ( c \cap ( ( ( a \cap c )
\cup ( b \cap d ) ) \cup
( ( ( a \cap e ) \cup ( b \cap f ) ) \cap
( ( c \cap e ) \cup ( d \cap f ) ) ) ) ) ) \le \nonumber \\
& & (a\to_1 k )\cap(((a'\to_1 k)\cup((c'\to_1 k )\cap
 (c{\buildrel e,k\over\equiv}a))))\,.
   \label{eq:6oaproof5}
\end{eqnarray}
We convert the 4OA law
$(c'\to_1 k) \cap (c'{\buildrel e,k\over\equiv}m')
\le (m'\to_1 k)$ to
$(c'\to_1 k) \cap (c{\buildrel e,k\over\equiv}m)
\le (m'\to_1 k)$ to
$m'\cap((m'\to_1 k)
\cup((c'\to_1 k)\cap(c{\buildrel e,k\over\equiv}m)))\le k$
using (\ref{eq:oalem9}).
We substitute $(a\to_1 k)'$ for $m$ and simplify with
(\ref{eq:oalem4}) and (\ref{eq:oalem5}) to obtain
$(a\to_1 k)\cap((a'\to_1 k)
\cup((c'\to_1 k)\cap(c{\buildrel e,k\over\equiv}a)))\le k$.
Combining with (\ref{eq:6oaproof5}) yields
\begin{eqnarray}
\lefteqn{ b \le a \to_1 k \qquad \&\qquad d \le c \to_1 k \qquad \&\qquad
 f \le e \to_1 k \qquad \Rightarrow} & & \nonumber \\
& b \cap ( a \cup ( c \cap ( ( ( a \cap c )
\cup ( b \cap d ) ) \cup
( ( ( a \cap e ) \cup ( b \cap f ) ) \cap
( ( c \cap e ) \cup ( d \cap f ) ) ) ) ) ) \le k\,.\nonumber\\
& &
   \label{eq:6oaproof6}
\end{eqnarray}
Letting
$k=(a\cap b)\cup(c\cap d)\cup(e\cap f)$ we have
\begin{eqnarray}
a'\le b & \Rightarrow & b\le a\to_1 k\label{eq:6oaproof1}\\
c'\le d & \Rightarrow & d\le c\to_1 k\label{eq:6oaproof2}\\
e'\le f & \Rightarrow & f\le e\to_1 k\label{eq:6oaproof3}
\end{eqnarray}
[e.g.~for (\ref{eq:6oaproof1}), using (\ref{eq:oalem8}) we
have $b\le a\to_1 b=a'\cup(a\cap(a\cap b))\le a'\cup(a\cap k)=a\to_1 k$]
from which we obtain (\ref{eq:6oadual}).

Conversely, assume
(\ref{eq:6oadual}) holds.  Let $a=g\to_1 k,\ b=g'\to_1 k,\ c=h\to_1 k,\
d=h'\to_1 k,\ e=j\to_1 k, f=j'\to_1 k$.  The hypotheses of
(\ref{eq:6oadual}) are satisfied using (\ref{eq:oalem3}).  Noticing
[with the help of (\ref{eq:oalem2})] that
the right-hand side of the resulting inequality is $\le k$, we have
$(g'\to_1 k)\cap((g\to_1 k) \cup((h\to_1 k)\cap(h{\buildrel
j,k\over\equiv}g)))\le k$, so $g\cap((g\to_1 k) \cup((h\to_1
k)\cap(h{\buildrel j,k\over\equiv}g)))\le k$.  Applying
(\ref{eq:oalem9}), we have the 4OA law $(h\to_1 k)\cap(h{\buildrel
j,k\over\equiv}g)\le g\to_1 k$.
\end{proof}

Thus we have demonstrated that the orthoarguesian law (\ref{eq:6oa}) can
be expressed by an equation with only 4 variables instead of 6. This is
in contrast to the stronger Arguesian law that has been shown by Haiman
to necessarily involve at least 6 variables.$\>$\cite{haiman}

\medskip

\begin{figure}[htbp]\centering
  \setlength{\unitlength}{1pt}
  \begin{picture}(260,150)(-10,-10)

    \put(0,13) { 
      \begin{picture}(124,120)(0,0) 
        \put(0,0){\line(0,1){60}}
        \put(68,0){\line(0,1){60}}
        \put(0,0){\line(1,1){34}}
        \put(0,30){\line(1,1){34}}
        \put(0,60){\line(1,1){34}}
        \put(68,0){\line(-1,1){34}}
        \put(68,30){\line(-1,1){34}}
        \put(68,60){\line(-1,1){34}}

        \put(0,0){\circle{10}}
        \put(0,30){\circle{10}}
        \put(0,60){\circle{10}}
        \put(68,0){\circle{10}}
        \put(68,30){\circle{10}}
        \put(68,60){\circle{10}}
        \put(17,17){\circle{10}}
        \put(17,47){\circle{10}}
        \put(17,77){\circle{10}}
        \put(51,17){\circle{10}}
        \put(51,47){\circle{10}}
        \put(51,77){\circle{10}}
        \put(34,34){\circle{10}}
        \put(34,64){\circle{10}}
        \put(34,94){\circle{10}}
      \end{picture}
    } 

    \put(180,0) { 
      \begin{picture}(124,120)(0,0) 
        \put(110.9,96.2){\line(-2,1){49.3}}
        \put(110.9,96.2){\line(1,-4){13}}
        \put(89,0){\line(4,5){35}}
        \put(89,0){\line(-1,0){54.8}}
        \put(34.2,0){\line(-4,5){35}}
        \put(12.2,96.2){\line(-1,-4){13}}
        \put(12.2,96.2){\line(2,1){49.3}}
        \put(5.5,70){\line(1,0){112}}
        \put(61.5,70){\line(0,-1){41.2}}
        \put(61.5,28.8){\line(6,-1){44.5}}

        \put(61.5,120.5){\circle{10}}
        \put(110.5,96.2){\circle{10}}
        \put(123.5,44){\circle{10}}
        \put(89,0){\circle{10}}
        \put(34,0){\circle{10}}
        \put(-0.5,44){\circle{10}}
        \put(12.5,96.2){\circle{10}}
        \put(86.2,108.4){\circle{10}}
        \put(117.5,70){\circle{10}}
        \put(106.0,21.4){\circle{10}}
        \put(61.5,0){\circle{10}}
        \put(17.1,21.4){\circle{10}}
        \put(5.5,70){\circle{10}}
        \put(36.9,108.4){\circle{10}}
        \put(61.5,70){\circle{10}}
        \put(61.5,49.2){\circle{10}}
        \put(61.5,28.8){\circle{10}}
        \put(83.8,25.1){\circle{10}}
      \end{picture}
    } 

  \end{picture}

  \caption{\hbox to3mm{\hfill}(a)
   Greechie diagram for $\widehat{\mbox{L}}$ from
   \protect\cite{go-gr}, Fig.~II;
  \hbox to2mm{\hfill} (b) Greechie diagram for L38.
\label{fig-Lhat-L38}}
\end{figure}
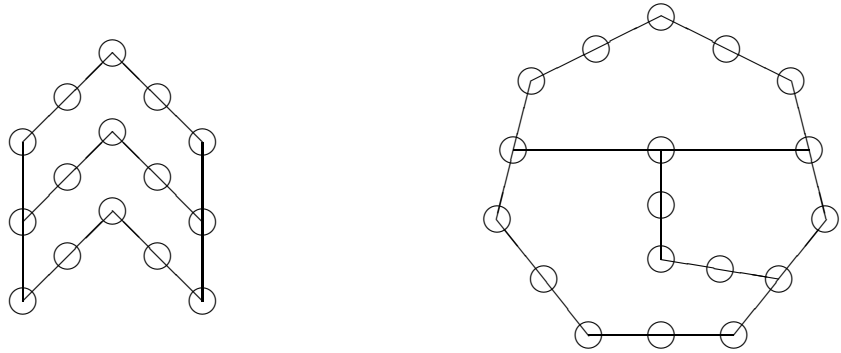

The 3OA law (\ref{eq:3oa}) expresses an orthoarguesian property that
does not hold in all OMLs, but as demonstrated by the fact that it holds
in OML L36, it is strictly weaker than the {\em proper} orthoarguesian
law expressed by (\ref{eq:4oa}) or (\ref{eq:6oa}).  The 3OA law is
equivalent to the following 3-variable equation \cite[Equation
III]{go-gr} obtained by Godowski and Greechie and thus to the other
3-variable variants of that equation mentioned in \cite{go-gr}.
Godowski and Greechie were apparently the first to observe that
(\ref{eq:go-gr3oa}) fails in OML L28 and also in OML
$\widehat{\mbox{L}}$ of Fig.~\ref{fig-Lhat-L38}a below.

\medskip\noindent
\begin{theorem}\label{th:go-gr3oa} An {\em OML} in which
\begin{eqnarray}
 \varphi_{b'}a\cup\alpha(a,b,c)=\varphi_{c'}a\cup\alpha(a,b,c)
 \label{eq:go-gr3oa}
\end{eqnarray}
[where $\varphi$ is the Sasaki projection
and $\alpha(a,b,c)\ {\buildrel\rm def\over =}
\ (b\cup c)\cap(\varphi_{b'}a\cup \varphi_{c'}a)$]
holds is a {\em 3OA} and vice versa.
\end{theorem}
\begin{proof}
Using the definitions, (\ref{eq:go-gr3oa}) can be written
in the dual form
$(a\to_1 c)\cap ((a\cap b)\cup((a\to_1 c)
\cap(b\to_1 c)))=
(b\to_1 c)\cap ((a\cap b)\cup((a\to_1 c)
\cap(b\to_1 c)))$.  We substitute $a'\to_1 c$ for $a$ and
$b'\to_1 c$ for $b$ throughout; simplifying with (\ref{eq:oalem4})
we obtain $(a\to_1 c) \cap (a{\buildrel c\over\equiv}b)=
(b\to_1 c) \cap (a{\buildrel c\over\equiv}b)$. This is easily
shown to be equivalent to (\ref{eq:3oa}).
\end{proof}

Equation (\ref{eq:go-gr3oa}) was derived by Godowski and Greechie from
Eq.~(\ref{eq:4oa-go-gr}) below, which is a 4-variable substitution
instance of (\ref{eq:6oa}).  Godowski and Greechie state that
(\ref{eq:go-gr3oa}) is ``more restrictive'' than (\ref{eq:4oa-go-gr}).
While it is not clear to us what is meant by this remark, it turns out
that the two equations are equivalent in an OML.  This equivalence also
means that the 4OA law cannot be derived from (\ref{eq:4oa-go-gr})
[which can also be verified independently by noticing that
(\ref{eq:4oa-go-gr}) does not fail in OML L36].

\medskip\noindent
\begin{theorem}\label{th:4oa-go-gr} An {\em OML} in which
\begin{eqnarray}
%
%
%
a \perp b \quad \&\quad c \perp d
\quad \Rightarrow \quad ( a \cup b ) \cap ( c \cup d ) \le
b \cup ( a \cap ( c \cup ( ( a \cup c )
\cap ( b \cup d ) ) ) ) \label{eq:4oa-go-gr}
\end{eqnarray}
holds is a {\em 3OA} and vice versa.
\end{theorem}
\begin{proof}
The proof is analogous to that for Theorem~\ref{th:6oa}.
\end{proof}

With the help of the following theorem we show that the relation of
equivalence introduced in Theorems \ref{th:oa-eq-c-3} and
\ref{th:oa-eq-c-4} is transitive.

\medskip\noindent
\begin{theorem}\label{th:oa-equiv} (a) In any {\em 3OA} we have:
\begin{eqnarray}
a{\buildrel c\over\equiv}b=1\qquad \Leftrightarrow\qquad
a\to_1c=b\to_1c
\label{eq:id1/c}
\end{eqnarray}
(b) In any {\em 4OA} we have:
\begin{eqnarray}
a{\buildrel c,d\over\equiv}b=1\qquad \Leftrightarrow\qquad
a\to_1d=b\to_1d
\label{eq:id1/cd}
\end{eqnarray}
\end{theorem}
\begin{proof}
For (\ref{eq:id1/c}), assuming $a{\buildrel c\over\equiv}b=1$ we have
$(a\to_1 c) \cap (a{\buildrel c\over\equiv}b)=(a\to_1 c)\cap 1
\le (b\to_1 c)$ by (\ref{eq:3oa}).  Conversely, from (\ref{eq:oalem3})
$(a\to_1 c)'\le a'\to_1 c$
and $(b\to_1 c)'\le b'\to_1 c$ and from the hypothesis
$(a\to_1 c)\equiv (b\to_1 c)=1$, so $1=((a\to_1 c)\cap(b\to_1 c))
\cup ((a\to_1 c)'\cap(b\to_1 c)')\le ((a\to_1 c)\cap(b\to_1 c))
\cup ((a'\to_1 c)\cap(b'\to_1 c))=a{\buildrel c\over\equiv}b$.
For (\ref{eq:id1/cd}), the proof is similar, noticing for its
converse that $a{\buildrel d\over\equiv}b\le
a{\buildrel c,d\over\equiv}b$.
\end{proof}

The inference rules (\ref{eq:id1/c}) and (\ref{eq:id1/cd}) fail in
lattices L28 and L36 respectively, suggesting the
possibility that they imply (in an OML) the 3OA and 4OA
laws.  However, we were unable to find a proof.

The transitive laws that are a consequence of (\ref{eq:id1/c}) and
(\ref{eq:id1/cd})
\begin{eqnarray}
a{\buildrel d\over\equiv}b=1\qquad \&\qquad
b{\buildrel d\over\equiv}c=1\qquad \Rightarrow\qquad
a{\buildrel d\over\equiv}c=1
\label{eq:tr1/c}\\
a{\buildrel d,e\over\equiv}b=1\qquad \&\qquad
b{\buildrel d,e\over\equiv}c=1\qquad \Rightarrow\qquad
a{\buildrel d,e\over\equiv}c=1
\label{eq:tr1/cd}
\end{eqnarray}
are weaker than the 3OA and 4OA laws, since both hold in lattice
$\widehat{\mbox{L}}$ of Fig.~\ref{fig-Lhat-L38}a (which violates both
laws). \begin{figure}[htbp]\centering
  \setlength{\unitlength}{1pt}
  \begin{picture}(260,150)(-10,-10)

    \put(0,0) { 
      \begin{picture}(124,120)(0,0) 
        \put(110.9,96.2){\line(-2,1){49.3}}
        \put(110.9,96.2){\line(1,-4){13}}
        \put(89,0){\line(4,5){35}}
        \put(89,0){\line(-1,0){54.8}}
        \put(34.2,0){\line(-4,5){35}}
        \put(12.2,96.2){\line(-1,-4){13}}
        \put(12.2,96.2){\line(2,1){49.3}}
        \qbezier(5.5,70)(61.5,35)(117.5,70)
        \qbezier(31,40)(61.5,65)(91,40)
        \put(36.9,108.4){\line(4,-5){69.4}}
        \put(86.2,108.4){\line(-4,-5){69.4}}
        \put(61.5,0){\line(0,1){52.5}}

        \put(61.5,120.5){\circle{10}}
        \put(110.5,96.2){\circle{10}}
        \put(123.5,44){\circle{10}}
        \put(89,0){\circle{10}}
        \put(34,0){\circle{10}}
        \put(-0.5,44){\circle{10}}
        \put(12.5,96.2){\circle{10}}
        \put(86.2,108.4){\circle{10}}
        \put(117.5,70){\circle{10}}
        \put(106.0,21.4){\circle{10}}
        \put(61.5,0){\circle{10}}
        \put(17.1,21.4){\circle{10}}
        \put(5.5,70){\circle{10}}
        \put(36.9,108.4){\circle{10}}
        \put(61.5,52.5){\circle{10}}
        \put(61.5,26.25){\circle{10}}
        \put(31,40){\circle{10}}
        \put(91,40){\circle{10}}
      \end{picture}
    } 

    \put(180,0) { 
      \begin{picture}(124,120)(0,0) 
        \put(110.9,96.2){\line(-2,1){49.3}}
        \put(110.9,96.2){\line(1,-4){13}}
        \put(89,0){\line(4,5){35}}
        \put(89,0){\line(-1,0){54.8}}
        \put(34.2,0){\line(-4,5){35}}
        \put(12.2,96.2){\line(-1,-4){13}}
        \put(12.2,96.2){\line(2,1){49.3}}
        \put(61.5,0){\line(0,1){96.2}}
        \put(12.5,96.2){\line(5,-4){93.5}}
        \put(110.5,96.2){\line(-5,-4){93.5}}
        \put(12.5,96.2){\line(1,0){49}}
        \put(110.5,96.2){\line(-5,-2){49}}

        \put(61.5,120.5){\circle{10}}
        \put(110.5,96.2){\circle{10}}
        \put(123.5,44){\circle{10}}
        \put(89,0){\circle{10}}
        \put(34,0){\circle{10}}
        \put(-0.5,44){\circle{10}}
        \put(12.5,96.2){\circle{10}}
        \put(86.2,108.4){\circle{10}}
        \put(117.5,70){\circle{10}}
        \put(106.0,21.4){\circle{10}}
        \put(61.5,0){\circle{10}}
        \put(17.1,21.4){\circle{10}}
        \put(5.5,70){\circle{10}}
        \put(36.9,108.4){\circle{10}}
        \put(61.5,96.2){\circle{10}}
        \put(61.5,76.8){\circle{10}}
        \put(42,41){\circle{10}}
        \put(81,41){\circle{10}}
        \put(40,96.2){\circle{10}}
        \put(86.2,86.5){\circle{10}}
      \end{picture}
    } 

  \end{picture}
  \caption{\hbox to3mm{\hfill}(a)
   Greechie diagram for OML L38m;
  \hbox to2mm{\hfill} (b) Greechie diagram for OML L42.
\label{fig:l42l38m}}
\end{figure}However, they have a weak orthoarguesian property:  both fail in
lattice L38m\footnote{OML L38m is neither a 3OA nor a 3GO, and in
addition violates all equations we have tested that are known not to
hold in all OMLs. It has been useful as a counterexample for
disproving equations conjectured to hold in all OMLs. OML L42 is a
4OA, a 5OA (Section \ref{sec:5oa}), and an $n$GO (for $n \le 9$, the
upper limit we have tested) but violates all equations we have tested
that are known to hold in neither 5OA nor 9GO.
It has been useful for disproving equations conjectured to hold in
at least one of these varieties.}
(Fig.~\ref{fig:l42l38m}a) and thus cannot be derived in an OML.

The 3OA law and its equivalents have been so far (to the authors'
knowledge) the only published 3-variable equations derived from the 4OA
law that do not hold in all OMLs.  Below we show another 3-variable
consequence of the 4OA law that is independent from the 3OA law.

%
%
%
%
%

\medskip\noindent
\begin{theorem}\label{th:new3oa} In any {\em OML}, the 3-variable equation
\begin{eqnarray}
(a\to_1 d) \cap (a{\buildrel c,d\over\equiv}a')
\le a'\to_1 d\,.\label{eq:new3oa}
\end{eqnarray}
holds in a {\rm 4OA} but cannot be derived from the {\em 3OA} law
nor vice versa.
\end{theorem}
\begin{proof}
This equation is obviously a substitution instance of the 4OA law
(\ref{eq:4oa}).  On the one hand, it fails in lattice L36 but holds in
lattice $\widehat{\mbox{L}}$ (Fig.~\ref{fig-Lhat-L38}a).  On the
other hand, the 3OA law
(\ref{eq:3oa}) holds in lattice L36 but fails in lattice
$\widehat{\mbox{L}}$.
\end{proof}

An interesting OML is L38 (Fig.~\ref{fig-Lhat-L38}b), which violates the
4OA law but does not violate any 3-variable consequence of the 4OA law
known to the authors.  One possibility that came to mind was that
perhaps L38 ``characterizes'' 4OA in an essential way, in the sense that
a failure in L38 of an equation derived in a 4OA implies its equivalence
to the 4OA law (analogous to the fact that a failure in O6 of an
equation derived in an OML implies its equivalence to the orthomodular
law).  This turns out not to be the case---there is a 4-variable
consequence of 4OA that is strictly weaker than 4OA but fails
in L38.  Whether there exists a 3-variable consequence of 4OA that
fails in L38 remains an open problem.

\medskip\noindent
\begin{theorem}\label{th:weak4oa}
A failure of a {\rm 4OA} equation in lattice {\rm L38} does not imply
its equivalence to the {\rm 4OA} law.
\end{theorem}
\begin{proof}
Writing the 4OA law as $(a'\to_1 d) \cap (a'{\buildrel c,d\over\equiv}b')
\le b'\to_1 d $, we weaken the left-hand side of the inequality
with $a\le(a'\to_1 d)$, etc.\ to obtain
\begin{eqnarray}
a\cap ((a\cap b)\cup
(((a\cap c)\cup (a'\cap (c\to_1 d)))\cap ((b\cap c)
\cup ((b\to_1 d)\cap c'))))\le b'\to_1 d \,.
   \label{eq:weak4oa}
\end{eqnarray}
This equation fails in OML L38 but holds in L28.
\end{proof}

L38 has a peculiar history. We found it ``by hand'' before we had
our present program for generating Greechie diagrams.
\cite{bdm-ndm-mp-1} Later we found out that G.~Beuttenm\"uller's
program \cite[pp.~319-328]{kalmb83} for generating Greechie diagrams
does not give L38 (and a number of other lattices). Looking for
a correct algorithm we came across McKay's isomorph-free
generation of graphs. Applied to Greechie diagrams it gave not
only a correct algorithm for their generation but also enabled
writing a program which is several orders of magnitude
faster than G.~Beuttenm\"uller's program transcribed into the C language
(originally it was written in Algol).

\section{\large Generalized Orthoarguesian Equations That Hold
in ${\cal C}({\cal H})$}
\label{sec:5oa}

Using the 3OA law as a starting point, we can construct an infinite
sequence of equations $E_1, E_2,\ldots$ that are valid in all Hilbert
lattices ${\cal C}({\cal H})$.  The second member $E_2$ of this sequence
is the 4OA law and the remaining members are equations with more
variables that imply the 4OA law.

\begin{definition}
\label{def:noa}
We define an operation
${\buildrel (n)\over\equiv}$ on $n$ variables
$a_1,\ldots,a_n$ ($n\ge 3$) as follows:\footnote{To obtain
${\buildrel (n)\over\equiv}$ we substitute in each
${\buildrel (n-1)\over\equiv}$ subexpression only the two explicit
variables, leaving the other variables the same.  For example,
$(a_2{\buildrel (4)\over\equiv}a_5)$ in (\ref{5oaoper}) means
$(a_2{\buildrel (3)\over\equiv}a_5)\cup ((a_2{\buildrel
(3)\over\equiv}a_4)\cap (a_5{\buildrel (3)\over\equiv}a_4))$ which means
$(((a_2\to_1 a_3)\cap(a_5\to_1 a_3)) \cup((a_2'\to_1 a_3)\cap(a_5'\to_1
a_3)))\cup
(
(((a_2\to_1 a_3)\cap(a_4\to_1 a_3))
\cup((a_2'\to_1 a_3)\cap(a_4'\to_1 a_3)))
\cap
(((a_5\to_1 a_3)\cap(a_4\to_1 a_3))
\cup((a_5'\to_1 a_3)\cap(a_4'\to_1 a_3)))
)
$.}
\begin{eqnarray}
a_1{\buildrel (3)\over\equiv}a_2\
&{\buildrel\rm def\over =}&\ a_1{\buildrel a_3\over\equiv}a_2\ =\
((a_1\to_1 a_3)\cap(a_2\to_1 a_3))
\cup((a_1'\to_1 a_3)\cap(a_2'\to_1 a_3))\quad \\
a_1{\buildrel (4)\over\equiv}a_2\
&{\buildrel\rm def\over =}&\ a_1{\buildrel a_4,a_3\over\equiv}a_2\
=\ (a_1{\buildrel (3)\over\equiv}a_2)\cup
((a_1{\buildrel (3)\over\equiv}a_4)\cap
(a_2{\buildrel (3)\over\equiv}a_4))\\
a_1{\buildrel (5)\over\equiv}a_2\
&{\buildrel\rm def\over =}&\ (a_1{\buildrel (4)\over\equiv}a_2)\cup
((a_1{\buildrel (4)\over\equiv}a_5)\cap
(a_2{\buildrel (4)\over\equiv}a_5))\label{5oaoper}\\
a_1{\buildrel (n)\over\equiv}a_2\
&{\buildrel\rm def\over =}&\ (a_1{\buildrel (n-1)\over\equiv}a_2)\cup
((a_1{\buildrel (n-1)\over\equiv}a_n)\cap
(a_2{\buildrel (n-1)\over\equiv}a_n))\,,\quad n\ge 4\,.
\end{eqnarray}
\end{definition}

Then we have the $n$OA {\em laws}
\begin{eqnarray}
(a_1\to_1 a_3) \cap (a_1{\buildrel (n)\over\equiv}a_2)
\le a_2\to_1 a_3\,.\label{eq:noa}
\end{eqnarray}

Each $n$OA law can be shown to be equivalent, in an OML, to equation
$E_{n-2}$ of Theorem~\ref{th:n-oa} below by a proof analogous to that
for Theorem~\ref{th:6oa}.  Thus they all hold in ${\cal C}({\cal H})$
and for $n\ge 4$ imply (in an OML) the 4OA law.  Also, as we shall show,
5OA is strictly smaller than 4OA, providing us with a new equational
variety valid in ${\cal C}({\cal H})$ that apparently has not been
previously known.  It remains an open problem whether in general, $n$OA
is strictly smaller than $(n-1)$OA.

For the following theorem we will refer to the 3OA and 4OA laws in their
4- and 6-variable forms (\ref{eq:4oa-go-gr}) and (\ref{eq:6oa}).
Starting with the 3OA law, we construct a sequence of equations as
follows.

\medskip\noindent
\begin{theorem}\label{th:n-oa}
Let $E_1, E_2,\ldots$ be the sequence of equations constructed as follows.
The first equation $E_1$ is the {\em 3OA} law expressed as
\begin{eqnarray}
%
\lefteqn{a_0 \perp b_0 \quad \&\quad a_1 \perp b_1
\quad \Rightarrow} \nonumber\\
& & ( a_0 \cup b_0 ) \cap ( a_1 \cup b_1 ) \le
b_0 \cup ( a_0 \cap ( a_1 \cup ( ( a_0 \cup a_1 )
\cap ( b_0 \cup b_1 ) ) ) )\,. \label{eq:n-oa1}
\end{eqnarray}
Given an equation $E_{n-1}$,
\begin{eqnarray}
\lefteqn{a_0 \perp b_0 \quad \&\quad a_1 \perp b_1
   \quad\ldots\quad\&\quad a_{n-1}\perp b_{n-1}
   \quad \Rightarrow} \nonumber \\
& & ( a_0 \cup b_0 ) \cap ( a_1 \cup b_1 )\cdots\cap (a_{n-1}\cup b_{n-1})
   \nonumber\\
& & \le b_0 \cup ( a_0 \cap ( a_1 \cup (
\cdots ( a_i \cup a_j ) \cap ( b_i \cup b_j )\cdots
 ) ) ) \label{eq:n-oa2}
\end{eqnarray}
we add new variables $a_n$ and $b_n$ to the hypotheses and left-hand
side of the conclusion.  For each subexpression appearing in the
right-hand side of the conclusion that is of the form $( a_i \cup a_j )
\cap ( b_i \cup b_j )$, $i,j<n$, we replace it with $( a_i \cup a_j )
\cap ( b_i \cup b_j )\cap((( a_i \cup a_n ) \cap ( b_i \cup b_n ))\cup((
a_j \cup a_n ) \cap ( b_j \cup b_n )))$ to result in equation $E_n$:
\begin{eqnarray}
\lefteqn{a_0 \perp b_0 \quad \&\quad a_1 \perp b_1
   \quad\ldots\quad\&\quad a_n\perp b_n
   \quad \Rightarrow} \nonumber \\
& & ( a_0 \cup b_0 ) \cap ( a_1 \cup b_1 )\cdots\cap (a_n\cup b_n)
   \nonumber\\
& & \le b_0 \cup ( a_0 \cap ( a_1 \cup (
\cdots ( a_i \cup a_j ) \cap ( b_i \cup b_j
))\nonumber\\
& & \qquad\cap((( a_i \cup a_n ) \cap ( b_i \cup b_n ))\cup(( a_j \cup
a_n ) \cap ( b_j \cup b_n )))\cdots
 ) ) )\,. \label{eq:n-oa3}
\end{eqnarray}
Then $E_2$ is the {\em 4OA} law, and $E_n$ $(n\ge 3)$ is an equation that
implies the {\em 4OA} law, that holds in ${\cal C}({\cal H})$, and
that cannot be inferred from {\em 4OA}.
\end{theorem}
\begin{proof}
It is obvious by definition that $E_2$ is the 4OA law (\ref{eq:6oa}).
It is also obvious $E_n$ $(n\ge 3)$ implies $E_{n-1}$ and thus the 4OA
law:  each subexpression of $E_{n-1}$ is greater than or equal to the
subexpression of $E_n$ that replaces it.

To show that $E_n$ holds in ${\cal C}({\cal H})$, we closely follow the
proof of the orthoarguesian equation in \cite{gr-non-s}.  We recall that
in lattice ${\cal C}({\cal H})$, the meet corresponds to set
intersection and $\le$ to $\subseteq$.  We replace the join with
subspace sum $\mbox{\boldmath $+$}$ throughout:  the orthogonality
hypotheses permit us to do this on the left-hand side of the conclusion
\cite[Lemma 3 on p.~67]{kalmb83}, and on the right-hand side we use
$a\mbox{\boldmath $+$}b\subseteq a\cup b$.

Suppose $x$ is a vector belonging to the left-hand side of
(\ref{eq:n-oa2}).  Then there exist vectors $x_0\in a_0,\ y_0\in b_0,\
\ldots ,\ x_{n-1}\in a_{n-1},\ y_{n-1}\in b_{n-1}$ such that
$x=x_0+y_0=\cdots=x_{n-1}+y_{n-1}$.  Hence $x_k-x_l=y_l-y_k$ for $0\le
k,l\le n-1$.  In (\ref{eq:n-oa2}) we assume, for our induction
hypothesis, that the components of vector $x=x_0+y_0$ can be distributed
over the leftmost terms on the right-hand side of the conclusion as
follows:
\[
  \cdots\subseteq
\underbrace{
  \underbrace{b_0}_{\textstyle y_0}
  \mbox{\boldmath $+$}(
  \underbrace{a_0}_{\textstyle x_0}
  \cap
  \underbrace{(
    \underbrace{a_1}_{\textstyle x_1}
    \mbox{\boldmath $+$}(
    \underbrace{
      \underbrace{(a_0\mbox{\boldmath $+$}a_1)}_{\textstyle x_0-x_1}
      \cap
      \underbrace{(b_0\mbox{\boldmath $+$}b_1)}_{\textstyle -y_0+y_1=x_0-x_1}
    }_{\textstyle x_0-x_1}
    \cap
    \underbrace{\cdots}_{\textstyle x_0-x_1}
    \cap
    \underbrace{\cdots}_{\textstyle x_0-x_1}
  }_{\textstyle x_1+(x_0-x_1)=x_0}
}_{\textstyle y_0+x_0=x}
\]
In particular if we eliminate the right-hand ellipses we obtain a ${\cal
C}({\cal H})$ proof of the starting equation $E_1$, which is the 3OA
law; this is the basis for our induction.

Let us first extend (\ref{eq:n-oa2}) by adding variables $a_n$ and $b_n$
to the hypothesis and left-hand side of the conclusion.  The extended
(\ref{eq:n-oa2}) so obtained obviously continues to hold in ${\cal
C}({\cal H})$.  Suppose $x$ is a vector belonging to the left-hand side
of this extended (\ref{eq:n-oa2}).  Then there exist vectors $x_0\in
a_0,\ y_0\in b_0,\ \ldots ,\ x_n\in a_n,\ y_n\in b_n$ such that
$x=x_0+y_0=\cdots=x_n+y_n$.  Hence $x_k-x_l=y_l-y_k$ for $0\le k,l\le
n$.  On the right-hand side of the extended (\ref{eq:n-oa2}), for any
arbitrary subexpression of the form $( a_i \cup a_j ) \cap ( b_i \cup
b_j )$, where $i,j<n$, the vector components will be distributed
(possibly with signs reversed) as $x_i-x_j \in
a_i\mbox{\boldmath $+$} a_j$ and $x_i-x_j=-y_i+y_j \in b_i\mbox{\boldmath $+$}
b_j$.  If we replace $( a_i \cup a_j ) \cap ( b_i \cup b_j )$ with $(
a_i \cup a_j ) \cap ( b_i \cup b_j )\cap((( a_i \cup a_n ) \cap ( b_i
\cup b_n ))\cup(( a_j \cup a_n ) \cap ( b_j \cup b_n )))$, components
$x_i$ and $x_j$ can be distributed as
\[
  \underbrace{( a_i \mbox{\boldmath $+$} a_j )}_{\textstyle x_i-x_j=}
  \cap
  \underbrace{( b_i \mbox{\boldmath $+$} b_j)}_{\textstyle -y_i+y_j}
  \cap
  \underbrace{
    ((
    \underbrace{( a_i \mbox{\boldmath $+$} a_n )}_{\textstyle x_i-x_n=}
    \cap
    \underbrace{( b_i \mbox{\boldmath $+$} b_n )}_{\textstyle -y_i+y_n})
    \mbox{\boldmath $+$}(
    \underbrace{( a_j \mbox{\boldmath $+$} a_n )}_{\textstyle -x_j+x_n=}
    \cap
    \underbrace{( b_j \mbox{\boldmath $+$} b_n )}_{\textstyle y_j-y_n}
    ))}_{\textstyle (x_i-x_n)+(-x_j+x_n)=x_i-x_j}
\]
so that $x_i-x_j$ remains an element of the replacement subexpression.
We continue to replace all subexpressions of the form $( a_i \cup a_j )
\cap ( b_i \cup b_j )$, where $i,j<n$, as above until they are
exhausted, obtaining (\ref{eq:n-oa3}).

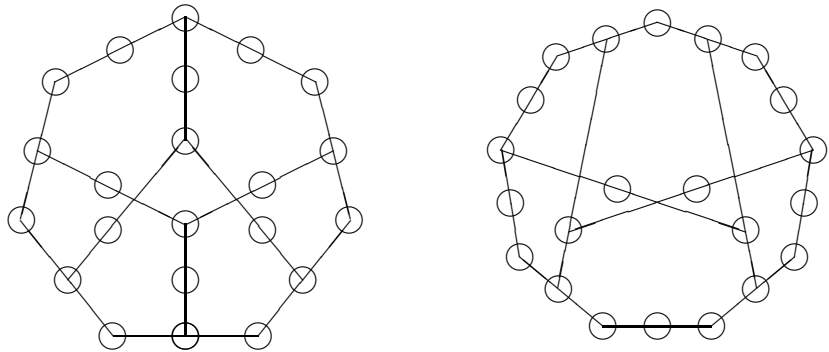
\begin{figure}[htbp]\centering
  \setlength{\unitlength}{1pt}
  \begin{picture}(260,150)(-10,-10)

    \put(0,0) { 
      \begin{picture}(124,120)(0,0) 

        \put(110.9,96.2){\line(-2,1){49.3}}
        \put(110.9,96.2){\line(1,-4){13}}
        \put(89,0){\line(4,5){35}}
        \put(89,0){\line(-1,0){54.8}}
        \put(34.2,0){\line(-4,5){35}}
        \put(12.2,96.2){\line(-1,-4){13}}
        \put(12.2,96.2){\line(2,1){49.3}}
        \put(61.5,74){\line(0,1){47}}
        \put(17.1,21.4){\line(5,6){44}}
        \put(106.0,21.4){\line(-5,6){44}}
        \put(5.5,70){\line(2,-1){56}}
        \put(117.5,70){\line(-2,-1){56}}
        \put(61.5,0){\line(0,1){42.5}}
        \put(61.5,120.5){\circle{10}}
        \put(110.5,96.2){\circle{10}}
        \put(123.5,44){\circle{10}}
        \put(89,0){\circle{10}}
        \put(34,0){\circle{10}}
        \put(-0.5,44){\circle{10}}
        \put(12.5,96.2){\circle{10}}
        \put(86.2,108.4){\circle{10}}
        \put(117.5,70){\circle{10}}
        \put(106.0,21.4){\circle{10}}
        \put(61.5,0){\circle{10}}
        \put(17.1,21.4){\circle{10}}
        \put(5.5,70){\circle{10}}
        \put(36.9,108.4){\circle{10}}
        \put(32.3,57){\circle{10}}
        \put(90.7,57){\circle{10}}
        \put(61.5,0){\circle{10}}
        \put(61.5,21.25){\circle{10}}
        \put(61.5,42.5){\circle{10}}
        \put(61.5,97.3){\circle{10}}
        \put(61.5,74){\circle{10}}
        \put(32.3,40){\circle{10}}
        \put(90.7,40){\circle{10}}
      \end{picture}
    }

    \put(180,0) { 
      \begin{picture}(124,110)(0,0) 
        \put(39.48,3.62){\line(1,0){41.04}}
        \put(39.48,3.62){\line(-6,5){31.44}}
        \put(80.52,3.62){\line(6,5){31.44}}
        \put(8.04,30){\line(-1,6){6.75}}
        \put(111.96,30){\line(1,6){6.75}}
        \put(0.91,70.42){\line(3,5){21.2}}
        \put(119.09,70.42){\line(-3,5){21.2}}
        \put(22.13,105.96){\line(3,1){38}}
        \put(97.87,105.96){\line(-3,1){38}}
        \put(78.9,112.3){\line(1,-5){18.8}}
        \put(41.1,112.3){\line(-1,-5){18.8}}
        \put(119.09,70.42){\line(-3,-1){92.5}}
        \put(0.91,70.42){\line(3,-1){92.5}}

        \put(80.52,3.62){\circle{10}}
        \put(39.48,3.62){\circle{10}}
        \put(111.96,30){\circle{10}}
        \put(8.04,30){\circle{10}}
        \put(119.09,70.42){\circle{10}}
        \put(0.91,70.42){\circle{10}}
        \put(97.2,105.8){\circle{10}}
        \put(21.8,105.8){\circle{10}}
        \put(60,118.6){\circle{10}}
        \put(60,3.62){\circle{10}}
        \put(97.24,17.81){\circle{10}}
        \put(22.76,17.81){\circle{10}}
        \put(115.525,50.21){\circle{10}}
        \put(4.475,50.21){\circle{10}}
        \put(107.83,89.19){\circle{10}}
        \put(12.17,89.19){\circle{10}}
        \put(79.285,112.5){\circle{10}}
        \put(40.715,112.5){\circle{10}}
        \put(45,55.7){\circle{10}}
        \put(75,55.7){\circle{10}}
        \put(26.5,40){\circle{10}}
        \put(93.5,40){\circle{10}}

      \end{picture}
    } 

  \end{picture}
  \caption{\hbox to3mm{\hfill}(a)
   Greechie diagram for L46-7;
  \hbox to5mm{\hfill} (b) Greechie diagram for L46-9.
\label{fig:5oa}}
\end{figure}

That $E_n\ (n\ge 3)$ cannot be inferred from the 4OA law follows from
the fact that the 4OA law holds in L46-7 (Fig.~\ref{fig:5oa})
while $(a_1\to_1 a_3) \cap (a_1{\buildrel (5)\over\equiv}a_2) \le
a_2\to_1 a_3$ fails in it. L46-9, Fig.~\ref{fig:5oa}, is the only
other lattice with this property among all Greechie
3-atoms-in-a-block lattices with 22 atoms and 13 blocks. L46-7 and
L46-9 are most probably the smallest Greechie
3-atoms-in-a-block lattices with that property: we scanned
some 80\%\ of smaller lattices and did not find any other.
\end{proof}

\begin{theorem}\label{th:noa-equiv}
In any {\em $n$OA} we have:
\begin{eqnarray}
a_1{\buildrel (n)\over\equiv}a_2=1\qquad \Leftrightarrow\qquad
a_1\to_1a_3=a_2\to_1a_3
\label{eq:noae}
\end{eqnarray}
This also means that $a_1{\buildrel (n)\over\equiv}a_2$ being
equal to one is a relation of equivalence.
\end{theorem}
\begin{proof} The proof is analogous to the proof of the Theorem
\ref{th:oa-equiv}.
\end{proof}

As with the Theorem \ref{th:oa-equiv} there is an open problem
whether $(a_1\to_1 a_3) \cap (a_1{\buildrel (n)\over\equiv}a_2)
\le a_2\to_1 a_3$ follow from Eq.~(\ref{eq:noae}). The fact that
Eq.~(\ref{eq:noae}) fails in L46-7 and L46-9 for n=5 indicates
that they might.

\section{\large Distributive Properties That Hold in ${\cal C}({\cal H})$}
\label{sec:distr}

The distributive law does not hold in either orthomodular or Hilbert
lattices, as it would make them become Boolean algebras; indeed the
failure of this law is the essential difference between these lattices
and Boolean algebras.  But by using the Godowski and orthoarguesian
equations that extend the orthomodular ones, we can also extend the
distributive properties of OMLs such as those provided by F-H.  These
can give us additional insight into the nature of the distributive
properties of Hilbert lattices, as well as provide us with additional
methods for further study of these lattices.
In this section we show several distributive properties that imply the
Godowski and orthoarguesian equations they are derived from, meaning
that they are the strongest possible in their particular form.

\begin{definition}\label{def:chained-equiv}
Let us call the following expression a {\em chained identity}:
\begin{eqnarray}
a_1{\buildrel\ldots\over\equiv}a_n{\buildrel{\rm def}
\over =}(a_1\equiv a_2)\cdots
\cap(a_{n-1}\equiv a_n),\qquad
n=2,3,4,\dots\label{eq:chained-equiv}
\end{eqnarray}
\end{definition}

\begin{lemma}\label{lem:chainedeq-com}
In any {\em OML\/}, the chained identity
$a_1{\buildrel\ldots\over\equiv}a_n$ commutes with every
term (polynomial) constructed from variables $a_1,\ldots,a_n$.
\end{lemma}
\begin{proof}
In any OML, from
(\ref{eq:om-alt}) we have
$(a\equiv b)\cap(b\equiv c)= (a\equiv b)\cap(a\equiv c)$.
Using this, we rewrite $a_1{\buildrel\ldots\over\equiv}a_n$ as
$(a_1\equiv a_2)\cdots\cap(a_1\equiv a_n)$.  In any OML we also have
$aCa\equiv b$.  Repeatedly applying the commutation law $aCb\ \&\ aCc
\Rightarrow aCb\cap c$, we prove
$a_1Ca_1{\buildrel\ldots\over\equiv}a_n$.  Similarly, for any
$1\le i\le n$ we have $a_iCa_1{\buildrel\ldots\over\equiv}a_n$.
Repeatedly applying the commutation laws $aCc\ \&\ bCc \Rightarrow a\cup
bCc$ and $aCb\Rightarrow a'Cb$, we can build up
$tCa_1{\buildrel\ldots\over\equiv}a_n$ for any expression $t$
constructed from variables $a_1,\ldots,a_n$.

As an exercise, the reader is invited to show an alternate proof using
(\ref{eq:god-prelemma3a}).
\end{proof}

\begin{theorem}\label{th:chainedgo-com}
In any {\em OML\/} in which the Godowski equation $n$-Go
holds, the Godowski identity $a_1{\buildrel\gamma\over\equiv}a_n$
commutes with any term constructed from variables
$a_1,\dots,a_n$.
\end{theorem}
\begin{proof}
Lemma \ref{lem:chainedeq-com} and (\ref{eq:godow1c}).
\end{proof}
We can use this commutation relationship in conjunction with
F-H to obtain immediately simple
distributive laws that hold for
any $n$GO, such as
$(a_1{\buildrel\gamma\over\equiv}a_n)\cap(s\cup t)=
((a_1{\buildrel\gamma\over\equiv}a_n)\cap s)\cup
((a_1{\buildrel\gamma\over\equiv}a_n)\cap t)$, where $s$ and $t$
are any terms constructed from variables $a_1,\ldots,a_n$.
More general laws are also possible, as Theorem \ref{th:godistr} below
shows.

\begin{lemma}\label{def:omldistr}
In any {\em OML} the following inferences hold.
\begin{eqnarray}
aCd\quad\&\quad bCd\quad\&\quad b\cap d\le c\le d\quad\Rightarrow\quad
  a\cap(b\cup c)=(a\cap b)\cup(a\cap c)
  \label{eq:omldistr1} \\
aCd\quad\&\quad bCd\quad\&\quad b\cap d\le c\le d\quad\Rightarrow\quad
  b\cap(a\cup c)=(b\cap a)\cup(b\cap c)
  \label{eq:omldistr2} \\
aCd\quad\&\quad c\le d\le b'\quad\Rightarrow\quad
  c\cap(a\cup b)=(c\cap a)\cup(c\cap b)
  \label{eq:omldistr3}
\end{eqnarray}
\end{lemma}
\begin{proof}
For (\ref{eq:omldistr1}):
$a\cap(b\cup c)
=a\cap(b\cup(c\cap d))
=$[using $bCd$, $cCd$] $a\cap((b\cup c)\cap(b\cup d))
=(b\cup c)\cap(a\cap(b\cup d))
=$[using $aCd$, $bCd$] $(b\cup c)\cap((a\cap b)\cup(a\cap d))
=$[using $b\cup cCa\cap b$, $a\cap bCa\cap d$]
  $((b\cup c)\cap(a\cap b))\cup((b\cup c)\cap(a\cap d))
=(a\cap b)\cup(a\cap(d\cap(b\cup c)))
=$[using $dCb$, $dCc$] $(a\cap b)\cup(a\cap((d\cap b)\cup(d\cap c))
=$[since $d\cap b=b\cap c$ and $c\le d$] $(a\cap b)
    \cup(a\cap((b\cap c)\cup c))
=(a\cap b)\cup(a\cap c)$.

For (\ref{eq:omldistr2}):
$b\cap(a\cup c)
\le(b\cap(a\cup d)
=$[using $bCd$, $aCd$] $(b\cap a)\cup(b\cap d)
=$[using $b\cap d=b\cap c$] $(b\cap a)\cup(b\cap c)$.  The other direction
of the inequality is obvious.

For (\ref{eq:omldistr3}):  $c\cap(a\cup b) =c\cap d\cap(a\cup b)
=$[using $dCa$, $dCb$] $c\cap((d\cap a)\cup(d\cap b)) =c\cap((d\cap
a)\cup 0) =(c\cap a)\cup 0 =(c\cap a)\cup(c\cap b)$.
\end{proof}
In passing we note that (\ref{eq:omldistr1})--(\ref{eq:omldistr3}) are
examples of OML distributive properties that cannot be obtained
directly from F-H because $a$ does not
necessarily commute with either $b$ or $c$ (lattice MO2 would violate
these conclusions).  Also, the conclusion of (\ref{eq:omldistr3}) does
not hold under the weaker hypotheses of (\ref{eq:omldistr1}) since the
inference would fail in OML L42 (Fig.~\ref{fig:l42l38m}b).    We also
mention that (\ref{eq:omldistr1})--(\ref{eq:omldistr3}) all fail in
lattice O6 and thus are equivalent to the orthomodular law.

The next theorem shows examples of more general distributive laws
equivalent to $n$-Go, where the variables $a$, $b$, and $c$ are not
necessarily equal to any other specific term and may be different from
variables $a_1,\ldots,a_n$.  The hypotheses of (\ref{eq:godistr1}) and
(\ref{eq:godistr2}) can also be replaced by those of (\ref{eq:godistr3})
to obtain simpler though somewhat less general laws.

\begin{definition}\label{def:chained-impl}
Let us call the following expression a {\em chained implication}:
\begin{eqnarray}
a_1{\buildrel{\ldots\ }\over\to}a_n{\buildrel{\rm def}
\over =}(a_1\to_1a_2)\cdots
\cap(a_{n-1}\to_1a_n),\qquad
n=2,3,4,\dots\label{eq:chained-impl}
\end{eqnarray}
\end{definition}

\begin{theorem}\label{th:godistr}
Let $t$ be any term constructed from variables $a_1,\ldots,a_n$.  Then
in any {\em $n$GO} ($n\ge 3$), we have the following distributive laws
for any variables $a,b,c$ not necessarily in the list $a_1,\ldots,a_n$:
\begin{eqnarray}
\lefteqn{a_1{\buildrel\ldots\over\equiv}a_n\le a\le
a_1{\buildrel{\ldots\ }\over\to}a_n
\quad\&\quad
bCt\quad\&\quad b\cap t\le c\le t\quad\&\quad
 b\cup c\le a_n\to_1 a_1
\quad\Rightarrow}
\qquad\qquad\qquad\qquad\qquad\qquad\qquad\qquad\qquad\qquad
 \nonumber \\
& & a\cap(b\cup c)=(a\cap b)\cup(a\cap c) \label{eq:godistr1} \\
\lefteqn{a_1{\buildrel\ldots\over\equiv}a_n\le a\le
a_1{\buildrel{\ldots\ }\over\to}a_n
\quad\&\quad
bCt\quad\&\quad b\cap t\le c\le t\quad\&\quad
 b\cup c\le a_n\to_1 a_1
\quad\Rightarrow}
\qquad\qquad\qquad\qquad\qquad\qquad\qquad\qquad\qquad\qquad
 \nonumber \\
& & b\cap(a\cup c)=(b\cap a)\cup(b\cap c) \label{eq:godistr2} \\
\lefteqn{a_1{\buildrel\ldots\over\equiv}a_n\le a\le
a_1{\buildrel{\ldots\ }\over\to}a_n
\quad\&\quad
c\le t\le b'\quad\&\quad
 b\cup c\le a_n\to_1 a_1
\quad\Rightarrow}
\qquad\qquad\qquad\qquad\qquad\qquad\qquad\qquad\qquad\qquad
 \nonumber \\
& & c\cap(a\cup b)=(c\cap a)\cup(c\cap b) \label{eq:godistr3}
\end{eqnarray}
In particular, when $t$ is $a_1\cap a_n$ [for (\ref{eq:godistr1})
and (\ref{eq:godistr2})] or $a_n'$ [for (\ref{eq:godistr3})],
an {\em OML} in which any one of these inferences holds is an {\em $n$GO}
and vice versa.
\end{theorem}
\begin{proof}
For (\ref{eq:godistr1}):
Let $s$ abbreviate $a\cap(a_n\to_1a_1)$.
By (\ref{eq:godow1c}) we have
$(a_1{\buildrel{\ldots\ }\over\to}a_n)\cap(a_n\to_1a_1)=
a_1{\buildrel\gamma\over\equiv}a_n=
a_1{\buildrel\ldots\over\equiv}a_n$.
Hence from the first hypothesis
$a_1{\buildrel\ldots\over\equiv}a_n
=(a_1{\buildrel\ldots\over\equiv}a_n)\cap(a_n\to_1a_1)
\le a\cap(a_n\to_1a_1)
\le(a_1{\buildrel{\ldots\ }\over\to}a_n)\cap(a_n\to_1a_1)
=a_1{\buildrel\ldots\over\equiv}a_n$,
so
$s=a\cap(a_n\to_1a_1)=a_1{\buildrel\ldots\over\equiv}a_n$
and by Lemma \ref{lem:chainedeq-com} $sCt$.
Using (\ref{eq:omldistr1}), we obtain
\begin{eqnarray}
\lefteqn{a_1{\buildrel\ldots\over\equiv}a_n\le a\le
a_1{\buildrel{\ldots\ }\over\to}a_n
\quad\&\quad
bCt\quad\&\quad b\cap t\le c\le t
\quad\Rightarrow}
\qquad\qquad\qquad\qquad\qquad\qquad\qquad\qquad\qquad\qquad
 \nonumber \\
& & s\cap(b\cup c)=(s\cap b)\cup(s\cap c)\,. \nonumber
\end{eqnarray}
Since $b\cup c\le a_n\to_1 a_1$, it follows that
$s\cap(b\cup c)=a\cap(b\cup c)$, $s\cap b=a\cap b$, and $s\cap c=a\cap c$.

In a similar way we obtain (\ref{eq:godistr2}) and (\ref{eq:godistr3})
from (\ref{eq:omldistr2}) and (\ref{eq:omldistr3}) respectively.

To obtain the $n$GO law from (\ref{eq:godistr1}),
we substitute $a_1{\buildrel\gamma\over\equiv}a_n$ for $a$,
$a_1\cap a_n$ for $t$ and $c$, and $a_n'$ for $b$.  The hypotheses
of (\ref{eq:godistr1}) are satisfied in any OML, and the conclusion becomes
$(a_1{\buildrel\gamma\over\equiv}a_n)\cap(a_n\to_1 a_1)
=((a_1{\buildrel\gamma\over\equiv}a_n)\cap a_n')\cup
((a_1{\buildrel\gamma\over\equiv}a_n)\cap(a_1\cap a_n))
=$[using (\ref{eq:goswap2})]
$((a_1{\buildrel\gamma\over\equiv}a_n)\cap a_1')\cup
((a_1{\buildrel\gamma\over\equiv}a_n)\cap(a_1\cap a_n))
\le a_1'\cup(a_1\cap a_n)$, which is (\ref{eq:godow1d}).

To obtain the $n$GO law from (\ref{eq:godistr2}),
we make the same substitutions as above.  The conclusion
becomes
$a_n'\cap((a_1{\buildrel\gamma\over\equiv}a_n)\cup(a_n\cap a_1))
=(a_n'\cap
   (a_1{\buildrel\gamma\over\equiv}a_n))\cup(a_n'\cap a_n\cap a_1)
=a_n'\cap(a_1{\buildrel\gamma\over\equiv}a_n)
=$[using (\ref{eq:goswap2})]
$a_1'\cap(a_1{\buildrel\gamma\over\equiv}a_n)
\le a_1'$.
Therefore
$(a_1\cap a_n)\cup
  ((a_n'\cap((a_1{\buildrel\gamma\over\equiv}a_n)\cup(a_n\cap a_1)))
\le (a_1\cap a_n)\cup a_1' = a_1\to a_n$.
The left-hand side evaluates as
$(a_1\cap a_n)\cup
  ((a_n'\cap((a_1{\buildrel\gamma\over\equiv}a_n)\cup(a_n\cap a_1)))
=((a_1\cap a_n)\cup a_n')
  \cap((a_1{\buildrel\gamma\over\equiv}a_n)\cup(a_n\cap a_1))
\ge ((a_1\cap a_n)\cup a_n')\cap(a_1{\buildrel\gamma\over\equiv}a_n)
=a_1{\buildrel\gamma\over\equiv}a_n$,
establishing (\ref{eq:godow1d}).

To obtain the $n$GO law from (\ref{eq:godistr3}),
we substitute $a_1{\buildrel\gamma\over\equiv}a_n$ for $a$,
$a_n'$ for $t$ and $c$, and $a_1\cap a_n$ for $b$.  After that
the proof is the same as for (\ref{eq:godistr2}).
\end{proof}

In a 3OA or 4OA we can also derive distributive properties that are
stronger than those that hold in OML.  In fact the laws we show in
Theorems \ref{th:dist3oa} and \ref{th:dist4oa} below are strong enough
to determine a 3OA or 4OA.  First we prove a technical lemma.

\begin{lemma}In any {\em OML} we have:
\begin{eqnarray}
&(a\to_1 c)\cap((a\to_1 c)\cap(b\to_1 c))'
  \cap(a'\to_1 c)\cap(b'\to_1 c)=0
  \label{eq:oalem12}\\
&(a\to_1 c)\cap (((a\to_1 c)
\cap(b\to_1 c))'
\to_i((a'\to_1 c)\cap(b'\to_1 c)))\le b\to_1 c,
\ i=1,2\ \ \label{eq:oalem13}
\end{eqnarray}
\end{lemma}
\begin{proof}
For (\ref{eq:oalem12}):  By F-H we have $d\cap e\cap c\cap((d\to_1
c)'\cup(e\to_1 c)')= (d\cap e\cap c\cap(d\to_1 c)')\cup (d\cap e\cap
c\cap(e\to_1 c)')=0\cup 0=0$.  {}From (\ref{eq:oalem1}) we have $d\cap
e\cap (d\to_1 c)=d\cap e\cap c$.  Combining these we have $d\cap
e\cap(d\to_1 c)\cap((d\to_1 c)'\cup(e\to_1 c)')=0$.  Substituting
$(a'\to c)$ for $d$ and $(b'\to c)$ for $e$ and simplifying with
(\ref{eq:oalem4}) gives the result.

For (\ref{eq:oalem13}), $i=1$: Expanding the definition of
$\to_i$ ($i=1$) and applying F-H, we have
$(a\to_1 c)\cap (((a\to_1 c)\cap(b\to_1 c))'
\to_i((a'\to_1 c)\cap(b'\to_1 c)))=
((a\to_1 c)\cap((a\to_1 c)\cap(b\to_1 c)))\cup
((a\to_1 c)\cap((a\to_1 c)\cap(b\to_1 c))'
  \cap(a'\to_1 c)\cap(b'\to_1 c))=$ [using (\ref{eq:oalem12})]
$((a\to_1 c)\cap(b\to_1 c))\cup 0\le (b\to_1 c)$.

For (\ref{eq:oalem13}), $i=2$: Expanding the definition of
$\to_2$ and applying F-H, we have
$(d\to_1 c)\cap (((d\to_1 c)\cap(e\to_1 c))'\to_2(d\cap e))=
((d\to_1 c)\cap d\cap e)\cup((d\to_1 c)\cap (e\to_1 c)\cap (d\cap e)')
=$ [using (\ref{eq:oalem1})] $(d\cap e\cap c)\cup
((d\to_1 c)\cap (e\to_1 c)\cap (d\cap e)')\le
(e'\cup(e\cap c))\cup(e\to_1 c)=e\to_1 c$.
Substituting $(a'\to c)$ for $d$ and $(b'\to c)$ for $e$
and simplifying with (\ref{eq:oalem4}) gives the result.
\end{proof}

\medskip\noindent
\begin{theorem}\label{th:dist3oa} An {\em OML} in which
\begin{eqnarray}
\lefteqn{
  d \le a \to_1 c
         \qquad \&\qquad  d \cap (b \to_1 c) \le e
         \qquad \&\qquad  e \cup f \le a{\buildrel c\over\equiv}b
         \qquad \Rightarrow} \hspace{18em}& & \nonumber \\
& &  d \cap (e \cup f) = (d \cap e) \cup (d \cap f)
 \label{eq:dist3oa}
\end{eqnarray}
holds is a {\em 3OA} and vice versa.
\end{theorem}
\begin{proof}
Assume that (\ref{eq:dist3oa}) holds.  Substitute
$a\to_1 c$ for $d$, $((a\to_1 c)\cap(b\to_1 c))'
\to_1((a'\to_1 c)\cap(b'\to_1 c))$ for $e$, and
$((a\to_1 c)\cap(b\to_1 c))'\to_2((a'\to_1 c)\cap(b'\to_1 c))$ for $f$.
It is easy to see the hypotheses of (\ref{eq:dist3oa}) are
satisfied [use (\ref{eq:oalem6}) to establish the third
hypothesis].  Using (\ref{eq:oalem6}), the left-hand side of the
conclusion evaluates to $(a\to_1 c) \cap (a{\buildrel c\over\equiv}b)$.
The right-hand side is $((a\to_1 c)\cap (((a\to_1 c)
\cap(b\to_1 c))'
\to_1((a'\to_1 c)\cap(b'\to_1 c))))\cup((a\to_1 c)\cap (((a\to_1 c)
\cap(b\to_1 c))'
\to_2((a'\to_1 c)\cap(b'\to_1 c))))$, which by (\ref{eq:oalem13})
is $\le(b\to_1 c)\cup(b\to_1 c)=b\to_1 c$, establishing the
3OA law (\ref{eq:3oa}).

Conversely, we show the 3OA law implies (\ref{eq:dist3oa}).  In any OML
we have from the third hypothesis $d\cap (e \cup f) \le d\cap
(a{\buildrel c\over\equiv}b)$.  {}From the first hypothesis and the 3OA
law (\ref{eq:3oa}) we obtain $d\cap a{\buildrel c\over\equiv}b \le
d\cap (b \to_1 c)$.  {}From the second hypothesis we have $d \cap (b \to_1
c) \le d\cap e \le (d\cap e)\cup (d\cap f)$.  Thus $ d \cap (e \cup f)
\le (d \cap e) \cup (d \cap f)$.  Since $ d \cap (e \cup f) \ge (d \cap
e) \cup (d \cap f)$ holds in any ortholattice, we conclude $ d \cap (e
\cup f) = (d \cap e) \cup (d \cap f)$.
\end{proof}

The proof of the 4OA version of this theorem shows an application of the
3OA distributive law (\ref{eq:dist3oa}), where we use it to construct
the inner terms of the 4OA equation.

\medskip\noindent
\begin{theorem}\label{th:dist4oa} An {\em OML} in which
\begin{eqnarray}
\lefteqn{e \le a \to_1 d
         \qquad \&\qquad   e \cap (b \to_1 d) \le f
         \qquad \&\qquad   f \cup g \le a{\buildrel c,d\over\equiv}b
         \qquad \Rightarrow} \hspace{18em}& & \nonumber \\
& &  e \cap (f \cup g) = (e \cap f) \cup (e \cap g)
 \label{eq:dist4oa}
\end{eqnarray}
holds is a {\em 4OA} and vice versa.
\end{theorem}
\begin{proof}
Assume that (\ref{eq:dist4oa}) holds.  Since $a{\buildrel
d\over\equiv}b\le a{\buildrel c,d\over\equiv}b$, we also have that
(\ref{eq:dist3oa}) holds.  So by Theorem \ref{th:dist3oa} we have
\begin{eqnarray}
(a\to_1 d) \cap (a{\buildrel d\over\equiv}b)
\le b\to_1 d\,.\label{eq:dist4oaproof1}
\end{eqnarray}
By Theorem \ref{th:dist3oa} we also have $(a\to_1 d) \cap (a{\buildrel
d\over\equiv}c) \le c\to_1 d$, so $(a\to_1 d) \cap (a{\buildrel
d\over\equiv}c) \cap (b{\buildrel d\over\equiv}c) \le (c\to_1 d)
\cap (b{\buildrel d\over\equiv}c)$; applying Theorem \ref{th:dist3oa}
again to the right-hand side we obtain
\begin{eqnarray}
(a\to_1 d) \cap (a{\buildrel d\over\equiv}c) \cap
(b{\buildrel d\over\equiv}c)
\le b\to_1 d\,.\label{eq:dist4oaproof2}
\end{eqnarray}
In (\ref{eq:dist4oa}) we substitute $a\to_1 d$ for $e$, $a{\buildrel
d\over\equiv}b$ for $f$, and $(a{\buildrel d\over\equiv}c)\cap(
b{\buildrel d\over\equiv}c)$ for $g$.  It is easy to see the hypotheses
of (\ref{eq:dist4oa}) are satisfied, and the conclusion gives us
\begin{eqnarray}
(a\to_1 d) \cap (a{\buildrel c,d\over\equiv}b)=
((a\to_1 d) \cap (a{\buildrel d\over\equiv}b))\cup(
(a\to_1 d) \cap (a{\buildrel d\over\equiv}c) \cap
(b{\buildrel d\over\equiv}c))\,.\label{eq:dist4oaproof3}
\end{eqnarray}
{}From (\ref{eq:dist4oaproof1}), (\ref{eq:dist4oaproof2}), and
(\ref{eq:dist4oaproof3}) we conclude the 4OA law (\ref{eq:4oa}).

For the converse, the proof that the 4OA law implies (\ref{eq:dist4oa})
is essentially identical to that for Theorem \ref{th:dist3oa}.
\end{proof}

\section{\large Conclusion}
\label{sec:conclusion}

Our investigation in the field of Hilbert lattices and therefore in
the field of Hilbert space and its subspaces in previous sections
resulted in several novel results and many decisive simplifications
and unifications of the previously known results mostly due to
our new algorithms for generation of Greechie lattices and automated
checking of Hilbert space equations and lattice equations in
general. So, the results have their own merit in the theory of
Hilbert space, quantum measurements, and the general lattice theory,
but as we stressed in the Introduction, we were prompted to attack
the problem of generating Hilbert lattice equations and their
possible connections with the quantum states (probability measures)
by recent development in the field of quantum computing. In
particular, we are interested in the problem of making a quantum
computer work as a quantum simulator. In order to enable this, we were
looking for a way to feed a quantum computer with an
algebra underlying a Hilbert space description of quantum systems.
Boolean algebra underlies any classical theory or model computed on
a classical computer and it imposes conditions (equations) on classical
bits \{0,1\} with the help of classical logic gates. For quantum theory
such an algebra is still not known. Quantum computation  at its present
stage manipulates quantum bits $\{|0\rangle,|1\rangle\}$ by means of
quantum logic gates (unitary operators) following algorithms for computing
particular problems. A general quantum algebra underlying Hilbert
space does exist, though. It is the Hilbert lattice we elaborated in
Section \ref{sec:states}. However, its present axiomatic definition
by means of universal and existential quantifiers and infinite
dimensionality does not allow us to feed a quantum computer with it.
What we would need is an equational formulation of the Hilbert
lattice. This would again contribute in turn to the theory of
Hilbert space subspaces which is poorly developed. It is significant
that there are two ways of reconstructing Hilbert space starting
from an ortholattice. One is a pure lattice one and it is presented
in Section \ref{sec:states}. The other one is a pure state one.
\cite{holl95} The equational approach unites them.

There were only two classes of such equations known so far: Godowski's
and Mayet's equations determined by the states defined on a lattice
and 4- and 6-variable orthoarguesian equations determined by the
projective geometry defined on it. To these ones, we add our newly
discovered---in Theorem \ref{th:n-oa}---generalized orthoarguesian
equations with $n$ variables. In order to interconnect and simplify
the already known results on the former equations and to obtain new
results we analyze the interconnections between an ortholattice and
states defined on it and obtain the following results.
\begin{itemize}
\item By Theorem \ref{th:god-eq} and Theorem \ref{th:strong-distr}
the difference between classical and quantum states is that there
is a single classical state for all lattice elements
while quantum states for different lattice elements are different;
\item By Theorem \ref{th:strong-distr} a classical state defined on
an ortholattice turns it into the Boolean algebra;
\item By Theorem \ref{th:god-eq} a strong state defined
on an ortholattice turns it into a variety smaller than OML in which
Godowski's equations hold.
\end{itemize}

On the other hand,
\begin{itemize}
\item by Theorem \ref{th:sol} there is a way of obtaining complex
infinite dimensional Hilbert\break space from the Hilbert lattice equipped
with several additional conditions and without invoking the notion of
state at all. States then follow by Gleason's theorem.

\end{itemize}

As for  Godowski's and Mayet's equations we obtain the following
results:
\begin{itemize}
\item Theorems~\ref{th:god-th1}, \ref{th:god-th2},
\ref{th:god-theorem3}, and
\ref{th:god-theorem4} present several
new results on and simplifications of Godowski's equations
based on the operation of identity given by the Definition
\ref{def:ident} which is also used to give a new formulation of
orthomodularity by Theorem \ref{th:other-eq};
\item New Greechie diagrams in which Godowski's equations with up to
seven variables fail are presented in Figures  \ref{fig-oag45m},
\ref{fig-oag6}, and \ref{fig-g7s}. They were obtained by a new
algorithm for generating Greechie diagrams and a new algorithm
for automated checking of passage
of lattice equations through them \cite{bdm-ndm-mp-1} (see footnote at
the end of Section \ref{sec:oml-eqs} and the comment at the end of
Section \ref{sec:oa}) and they are by several atoms and blocks smaller
than the previously known ones. This makes preliminary checking of any
conjecture related to Godowski's equations a lot faster;
\item In Theorem \ref{th:mayet} Mayet's examples which were apparently
supposed to differ from Godowski's equations are derived from
Godowski's equations.
\end{itemize}

As for the orthoarguesian equations, their consequences,
and generalizations, the clue for their unification were 3- and
4-variable orthoarguesian identities (3-oa and 4-oa, defined in
Definition \ref{def:3-4-oa}) which served us to obtain the following
results:
\begin{itemize}
\item A 4-variable Eq.~(\ref{eq:4oa}), the 4OA law, is equivalent to
the original 6-variable orthoarguesian equation as given by
A.~Day as we show in Theorem \ref{th:6oa};
\item All lower than
6-variable consequences of the original orthoarguesian equation that one
can find in the literature can be reduced to the 3-variable
Eq.~(\ref{eq:3oa}), the 3OA law, as illustrated by Theorems
\ref{th:go-gr3oa} and \ref{th:4oa-go-gr};
\item There is a 3-variable consequence of
the 4OA law which is not equivalent to the 3OA law as proved in
Theorem \ref{th:new3oa};
\item There is an $n$-variable generalization of the orthoarguesian
equations, the $n$OA law which holds in any Hilbert lattice, as proved in
Theorem \ref{th:n-oa}, and which cannot be derived from the 4OA law,
as proved in Theorem \ref{th:noa-equiv};
\item The $n$OA law added to an ortholattice turns it into a variety
smaller than OML as shown by Theorems
\ref{th:oa-equiv}, \ref{th:oa-eq-c-3}, and \ref{th:oa-eq-c-4};
\item Each $n$OA determines a relation of equivalence as proved by
Theorem \ref{th:noa-equiv}.
\end{itemize}

In the end, different distributive properties that hold in the lattice of
closed subspaces of any Hilbert space are given in Section
\ref{sec:distr} and several intriguing open problems are formulated
following Theorems \ref{th:ident-distr-eq}, \ref{th:god-theorem3},
\ref{th:god-theorem4}, \ref{th:new3go}, \ref{th:oa-eq-c-4},
\ref{th:oa-equiv}, \ref{th:new3oa}, and \ref{th:noa-equiv}, as well as
preceding Theorems \ref{th:mayet} and \ref{th:god-trans}. Open problems
are also to attach a geometric interpretation to $n$OA and to
rigorously prove that infinite dimensional Hilbert space contains an
infinite sequence of relations of equivalence. The latter claim would
immediately follow from condition 5 of Theorem \ref{th:sol}, if
we could prove that for no $n$ can the $n$OA law can be inferred from
the $(n-1)$OA law starting with the $n=5$ case proved in Theorem
\ref{th:n-oa}. \cite[p.~379]{gr-non-s}

\parindent=20pt
\bigskip

\bigskip
\bibliographystyle{report}

\end{document}